%% file: NIKA_CL1227.tex
\documentclass[twocolumn,traditabstract]{aa}  

\usepackage{fixltx2e}
\usepackage[english]{babel}
\usepackage{graphicx,amsmath}
\usepackage{epstopdf}
\usepackage{epsf,color}
\usepackage[mathscr]{eucal}
\usepackage{amsmath}
\usepackage{amssymb,amsfonts}
\usepackage{natbib}
\usepackage{graphicx}
\usepackage{txfonts}
\usepackage{dsfont}
\definecolor{Mygreen}{rgb}{0.00, 0.72, 0.0}
\definecolor{Mypink}{rgb}{1.0, 0.0, 0.5}
\usepackage[breaklinks, citecolor=blue, linkcolor=Mygreen, urlcolor=Mypink, colorlinks=true, debug, baseurl=' ']{hyperref}
\usepackage{float} 
\usepackage{color}

\def\simlt{\lower.5ex\hbox{$\; \buildrel < \over \sim \;$}}
\def\simgt{\lower.5ex\hbox{$\; \buildrel > \over \sim \;$}}

\bibpunct{(}{)}{;}{a}{}{,} 
\begin{document}
%###############################################################################################
%##########################           START THE PAPER           ##########################################
%###############################################################################################
\title{Pressure distribution of the high-redshift cluster of galaxies \mbox{CL~J1226.9+3332} with NIKA \thanks{FITS file of the published maps are only available at the CDS via anonymous ftp to cdsarc.u-strasbg.fr (130.79.128.5) or via \url{http://cdsarc.u-strasbg.fr/viz-bin/qcat?J/A+A/576/12}}}
\input{listeauthors}
\date{Received \today \ / Accepted --}

\abstract {The thermal Sunyaev-Zel'dovich (tSZ) effect is expected to provide a low scatter mass proxy for galaxy clusters since it is directly proportional to the cluster thermal energy. The tSZ observations have proven to be a powerful tool for detecting and studying them, but high angular resolution observations are now needed to push their investigation to a higher redshift. In this paper, we report high angular (\textless 20 arcsec) resolution tSZ observations of the high-redshift cluster \mbox{CL~J1226.9+3332} ($z=0.89$). It was imaged at 150 and 260~GHz using the NIKA camera at the IRAM 30-meter telescope. The 150~GHz map shows that \mbox{CL~J1226.9+3332} is morphologically relaxed on large scales with evidence of a disturbed core, while the 260~GHz channel is used mostly to identify point source contamination. NIKA data are combined with those of Planck and \mbox{X-ray} from Chandra to infer the cluster's radial pressure, density, temperature, and entropy distributions. The total mass profile of the cluster is derived, and we find $M_{500} = 5.96^{+1.02}_{-0.79} \times 10^{14} M_{\sun}$ within the radius $R_{500} = 930^{+50}_{-43}$ kpc, at a 68\% confidence level. ($R_{500}$ is the radius within which the average density is 500 times the critical density at the cluster's redshift.) NIKA is the prototype camera of NIKA2, a KIDs (kinetic inductance detectors) based instrument to be installed at the end of 2015. This work is, therefore, part of a pilot study aiming at optimizing tSZ NIKA2 large programs.}

\titlerunning{NIKA observations of CL~J1226.9+3332}
\authorrunning{R. Adam, B. Comis, J.-F. Mac\'ias-P\'erez et al.}
\maketitle
%\tableofcontents

%###############################################################################################
%##########################                             INTRODUCTION                              ##########################%###############################################################################################
\section{Introduction}\label{sec:introduction}
%---------- Galaxy clusters for cosmology
Galaxy clusters are the largest gravitationally bound objects in the Universe. They arise from the collapse of primordial matter fluctuations, forming overdensity peaks at the intersection of filamentary structures. They offer a unique tracer of the matter distribution and a powerful probe for cosmology because they form across the expansion of the Universe. (See, for example, \cite{allen2011} and references therein for a detailed review.)

%---------- Cluster observable and cross
Clusters are mainly made of dark matter (about 85\% of their total masses), but also of hot ionized gas (about 12\%) and of the stars and interstellar medium within galaxies (a few percent), representing the baryonic component that can be used to detect and study them. Optical observations have been historically used to measure their total mass \citep{zwicky1933} -- typically around $10^{14} M_{\sun}$ -- from galaxy velocity dispersion and, more recently, from lensing measurements of background objects \citep[see][for a review]{bartelmann2010}. Since galaxy clusters form by accreting of surrounding material (dark matter, galaxies, and gas) and by merging with other clusters, they can be the source of a significant amount of non-thermal emission. Radio measurements around 1 GHz are used to explore such processes \citep[e.g.,][]{feretti2011}. The hot gas contained in the intracluster medium (ICM) -- a few keV -- emits X-ray photons due to the bremsstrahlung of energetic electrons \citep[see][]{sarazin1988}. Therefore, X-ray imaging can be used to study the electronic density distribution in galaxy clusters (with a weak dependence on the temperature). In addition, X-ray spectroscopy provides the possibility of measuring the ICM temperature \citep[see, for example,][]{bohringer2010}. 

%---------- The SZ as a complementary probe
To be used for cosmology, galaxy clusters observables need to be related in some way to their total mass. The precise calibration of such scaling relations requires to use as many available probes (that are complementary to each other) as possible. The thermal Sunyaev-Zel'dovich \citep[tSZ,][]{sunyaev1972,sunyaev1980} effect provides such a probe. It is due to the inverse Compton scattering of cosmic microwave background (CMB) photons with high-energy electrons in the ICM. The photons are shifted to higher frequencies providing a characteristic spectral distortion of the CMB, observable at millimeter wavelengths. Since the observable is not the cluster itself but the CMB, tSZ offers a key advantage because it does not suffer from cosmological dimming as do other probes. Its amplitude is directly proportional to the pressure distribution in clusters and is therefore expected to provide a low scatter mass proxy when assuming hydrostatic equilibrium \citep[e.g.,][]{nagai2006}. Together with X-ray observations, the tSZ effect allows for a detailed characterization of the ICM thermodynamics. See \cite{birkinshaw1999}, \cite{carlstrom2002}, and \cite{kitayama2014} for detailed reviews on the tSZ effect.

%---------- Importance of high z after Planck
The Planck satellite \citep{planck2013catalogue}, the South Pole Telescope \citep[SPT,][]{reichardt2013,bleem2014}, and the Atacama Cosmology Telescope \citep[ACT,][]{hasselfield2013} have produced, and will continue to improve, large tSZ selected cluster samples. However, as the high-redshift end of these samples is reached, clusters are not resolved owing to insufficient available angular resolution (larger than 1 arcmin). High angular resolution follow-up observations of these objects are needed to precisely calibrate the tSZ cluster observable versus their total mass, through their pressure profiles. The universality of such pressure distributions \citep{planck2013pressure_profile, arnaud2010}, taken as a standard candle, also has to be tested against redshift.

%---------- Previous observations of CLJ1226.9+3332
The object \mbox{CL~J1226.9+3332} is a high-redshift, hot and massive cluster of galaxies at $z = 0.89$. It was discovered in the WARPS survey  \citep[Wide Angle ROSAT Pointed Survey,][]{ebeling2001} and has been the object of multiwavelength studies. Owing to difficulty of \mbox{X-ray} spectroscopy at high-redshift, the first temperature estimates were made from SZA (Sunyaev-Zel'dovich Array) observations \citep[][$T_e~=~9.8^{+4.7}_{-1.9}$ keV]{joy2001}, providing the first confirmation that it is indeed a massive system. A detailed \mbox{X-ray} analysis of \mbox{XMM-Newton} observations by \cite{maughan2004} reports a consistent temperature, $T_e~=~11.5^{+1.1}_{-0.9}$ keV. They also measured \mbox{CL~J1226.9+3332} to show evidence of a relaxed \mbox{X-ray} morphology, in agreement with ROSAT first observations, and provided a total mass of $\left(1.4~\pm~0.5\right)~\times~10^{15} \ M_{\sun}$. More recent Chandra observations also agree that \mbox{CL~J1226.9+3332} is a hot system \citep[][$T_e~=~14.0^{+2.1}_{-1.8}$ keV]{bonamente2006}. The pressure profile of the cluster was measured at arcmin angular scales using the interferometric SZA observations at 30 and 90~GHz \citep{muchovej2007,mroczkowski2009,mroczkowski2011}, providing a detailed picture of the ICM on these scales. First indications of a disturbed core were made by an XMM/Chandra analysis, showing an asymmetry in the temperature map with a hotter southwest region \citep{maughan2007}. Lensing observations by the Hubble Space Telescope \citep[HST][]{jee2009} found a relaxed morphology on large scales and agree on the presence of a disturbed core on smaller scales, with the presence of a subclump 40 arcsec toward the southwest. This is consistent with the hotter region and highly correlated with the cluster galaxy distribution. They propose a scenario in which a less massive system has passed through the main cluster and the gas has been stripped during this passage. The mass inferred within the radius\footnote{The radius $r_{\Delta}$ corresponds to a radius within which the mean cluster density is $\Delta$ times the one of the critical density of the Universe at that redshift. Hereafter, we commonly use the physical quantities within this radius, noted with a subscript $\Delta$ generally taken as 500.} $r_{\Delta=200}$ is $(1.4 \pm 0.2) \times 10^{15} M_{\sun}$. Finally, MUSTANG tSZ observations at 90~GHz on $\sim$ 8 -- 45 arcsec scales \citep{korngut2011} have revealed a narrow 20 arcsec long ridge 10 arcsec southwest of the \mbox{X-ray} peak, in addition to another peak coincident with \mbox{X-ray} and the brightest cluster galaxy.

%---------- What we do here
In this paper, we report 150 and 260~GHz tSZ observations of \mbox{CL~J1226.9+3332}, using the New IRAM KIDs Array \citep[NIKA, see][for more details on the NIKA camera]{monfardini2010,bourion2011,monfardini2011,calvo2012,adam2013,catalano2014} at the IRAM (Institut de Radio Astronomie Millim\'etrique) 30-meter telescope. The reconstructed tSZ map of the cluster is used to constrain its pressure distribution, as well as the thermodynamics of the ICM gas by combining it with X-ray data. Since NIKA is the prototype of the future NIKA2 camera, these observations are part of a pilot study that aims at showing the potential of NIKA2 for follow-ups of unresolved Planck and ACT clusters. 

%---------- Paper organization
The paper is organized as follows. The observations of \mbox{CL~J1226.9+3332} are presented in Sect.~\ref{sec:CL1227Obs}, including the reduction of NIKA data. In Sect.~\ref{sec:analysis}, we present the analysis performed to recover the thermodynamical properties of the cluster. The results are provided in Sect.~\ref{sec:results} and compared to {data of other observatories} and previous observations. The conclusions and NIKA2 perspectives are given in Sect.~\ref{sec:conclusion}. Throughout this paper we assume a flat $\Lambda$CDM cosmology according to the latest {\it Planck} results \citep{planck2013param} with $H_0 = 67.11$ km s$^{-1}$ Mpc$^{-1}$, $\Omega_M = 0.3175$, and $\Omega_{\Lambda} = 0.6825$.

%###############################################################################################
%##########################                         NIKA OBSERVATIONS                         ##########################%###############################################################################################
\section{High resolution thermal Sunyaev-Zel'dovich observations}\label{sec:CL1227Obs}
%========== Observation with NIKA
\subsection{Observations of \mbox{CL~J1226.9+3332}}\label{sec:observations}
	\begin{figure*}[h]
	\centering
	\includegraphics[width=0.48\textwidth]{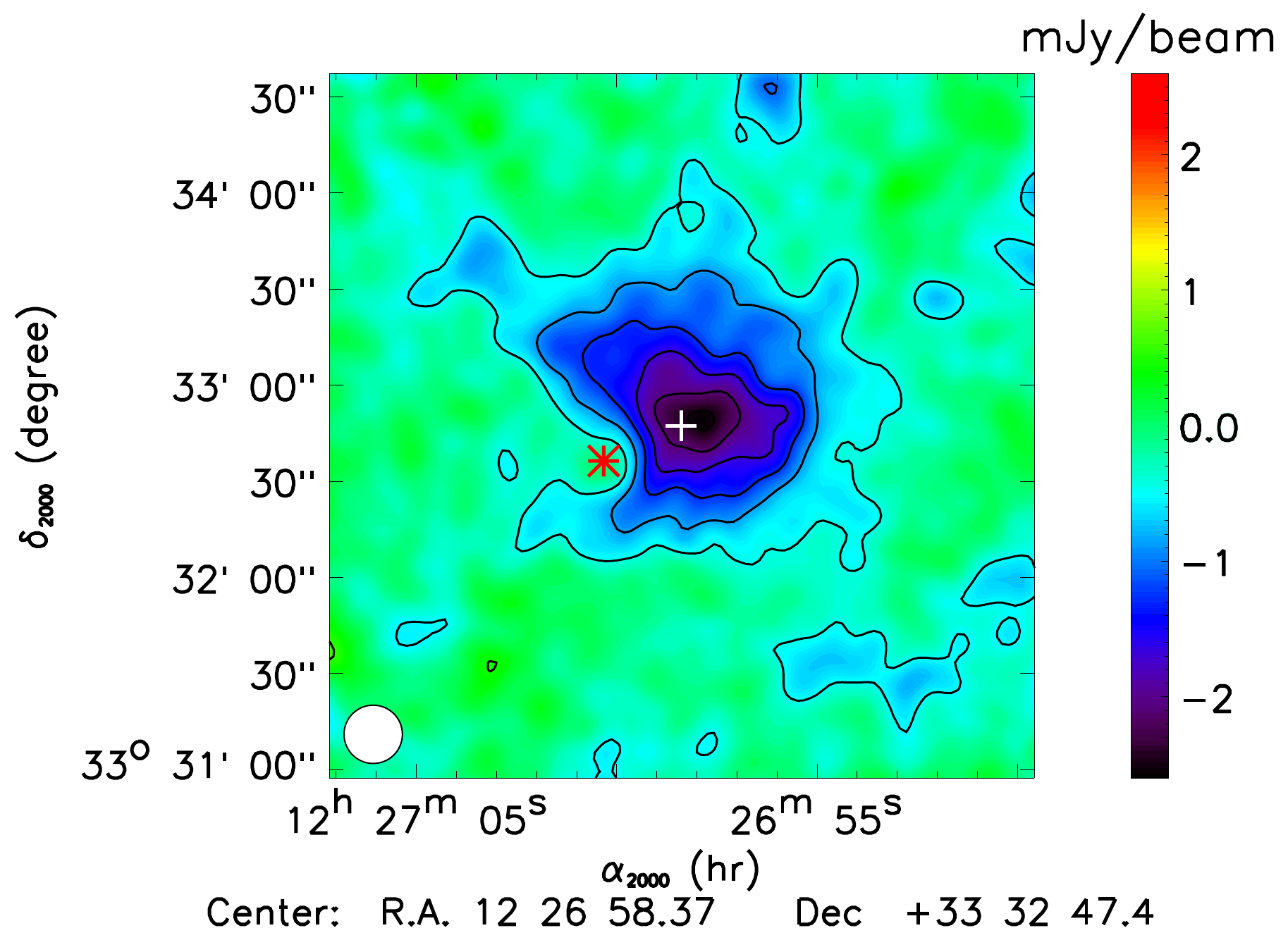}
	\includegraphics[width=0.48\textwidth]{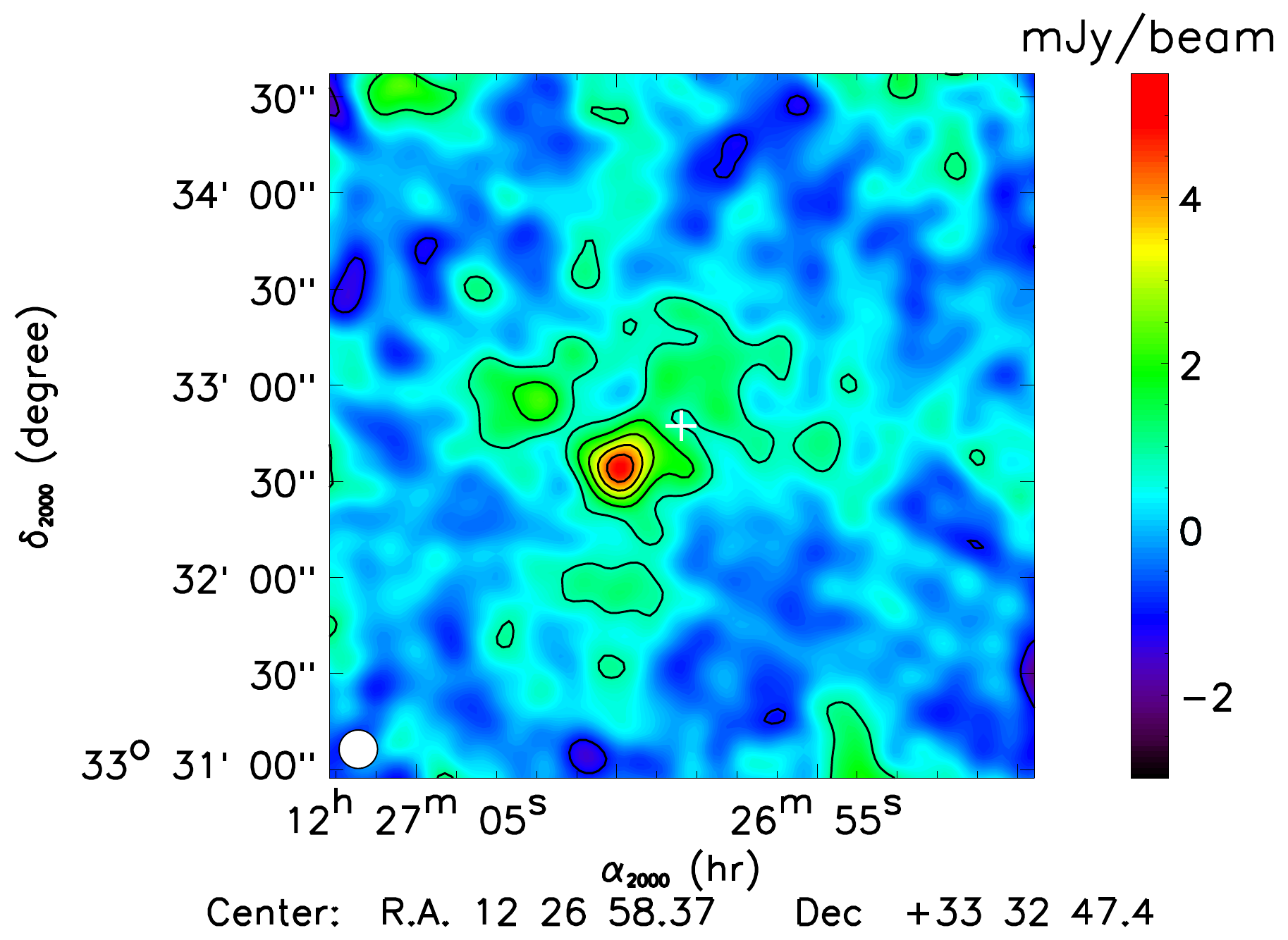}
	\caption{NIKA raw maps of \mbox{CL~J1226.9+3332} at 150~GHz (left) and 260~GHz (right). Contours are multiples of 3$\sigma$, excluding the zero level, which is not shown. The effective beams FWHM (12.0 and 18.2 arcsecond native resolution) are shown as the bottom left white circles, although the display images are smoothed with an extra 10 arcsec FWHM Gaussian. The position of the \mbox{X-ray} center is shown as a white cross in both maps and that of the point source is shown as a red star in the 150~GHz map.}
        \label{fig:CL1226map}
	\end{figure*}
	
	\begin{figure}[h]
	\centering
	\includegraphics[width=0.48\textwidth]{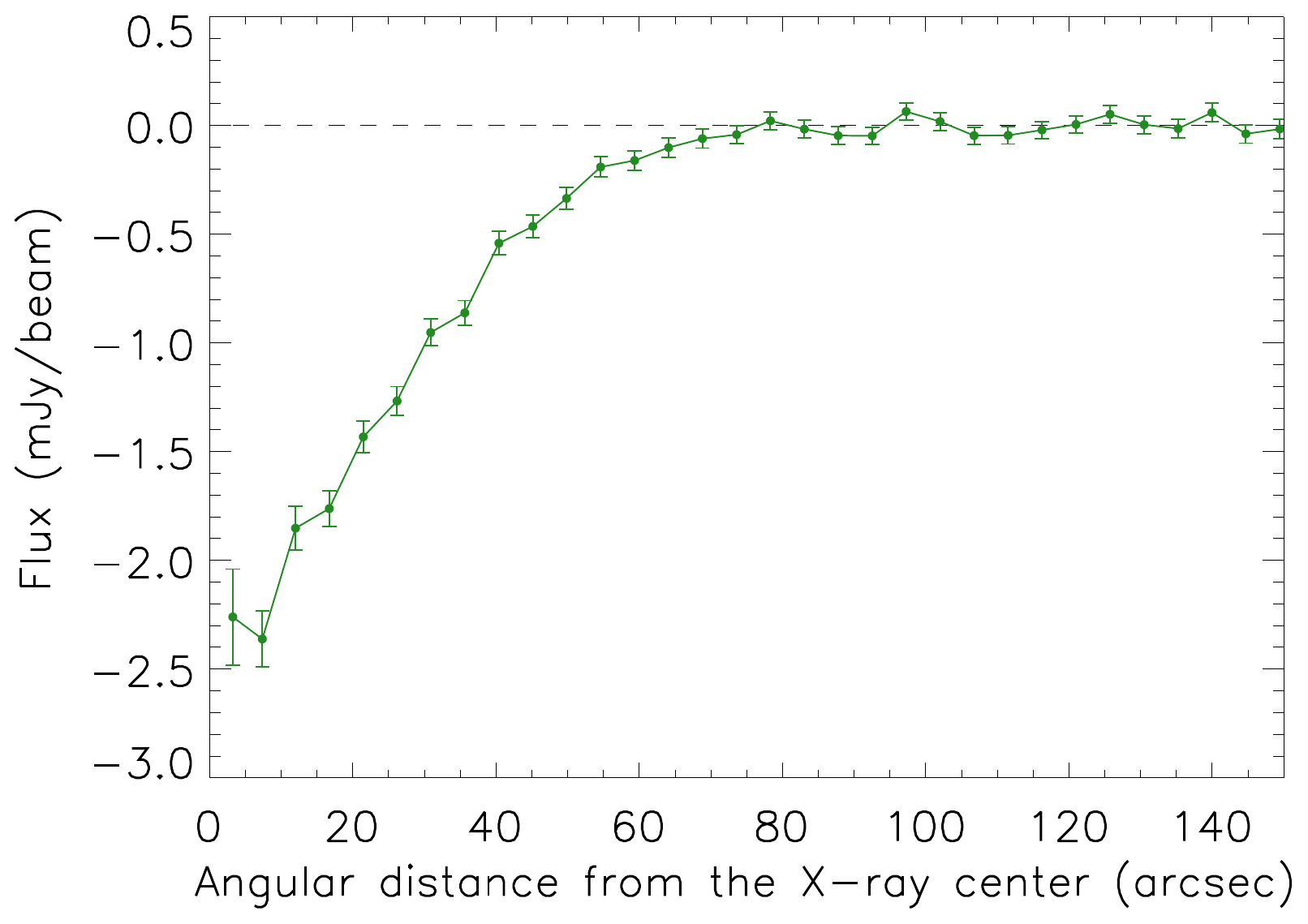}
	\caption{Flux density profile of \mbox{CL~J1226.9+3332} as measured by NIKA at 150~GHz, {\it i.e.}, the radial average within concentric annuli. The point source has been fitted and subtracted (see~Sect.~\ref{sec:param_estim}) before computing the profile. Error bars are only statistical.}
        \label{fig:CL1226profile2mm}
	\end{figure}
	
%---------- Scanning strategy
The NIKA camera was used at the IRAM 30-meter telescope (Pico Veleta, Spain) to image \mbox{CL~J1226.9+3332} at 150 and 260~GHz during the first NIKA open pool of February 2014. The cluster was mapped using on-the-fly raster scans made of constant azimuth -- resp. elevation -- subscans. Scans were made of 19 subscans of 6 arcmin length, separated by 10 arcsec elevation -- resp. azimuth -- steps. The subscan duration was fixed to ten seconds, giving a scanning speed of 36 arcsec per second and a total time of 3.3 minutes per scan. The pointing center was chosen to be (R.A.,~Dec.)~=~(12h~26m~58s,~33$^o$~32'~40") based on MUSTANG tSZ observations \citep{korngut2011}. All coordinates in this paper are given in equinox 2000.

%---------- Observing time and condition
The data collected were taken with an opacity, at 150~GHz (respectively 260~GHz), in the range 0.06--0.23 (respectively 0.06--0.29) and a mean value of 0.13 (respectively 0.16), corresponding to average winter conditions. The observations were mostly carried out during night time. A small fraction of the scans were flagged due to bad weather conditions and some others were lost because of missing data streams with the telescope position. The overall effective observing time on the cluster is 7.8 hours.

%---------- Pointing, focus, calibration, opacity
The overall final pointing residual errors were obtained with a precision of less than 3 arcsec using the observations of nearby quasars, 1308+326 and 1156+295, every hour. Uranus was taken as our primary calibrator, and we used its frequency-dependent brightness temperature model as given by \cite{moreno2010}, assumed to be accurate at the level of 5\%, as shown by \cite{planck2013calib}. Within the NIKA bandpasses of the February 2014 campaign, we obtain a mean brightness temperature of 112.7 and 92.8~K at 150 and 260~GHz, respectively. Uncertainties on the calibration were measured to be 5\% and 11\% at 150 and 260~GHz, respectively, using the dispersion of the recovered flux on Uranus maps. This corresponds to 7\% and 12\% overall calibration uncertainties when including the model error. The focus of the telescope was checked on Uranus or other bright point sources every two to three hours and systematically after sunset and sunrise. The effective FWHM was measured to be 18.2 and 12.0 arcsec at 150 and 260~GHz, respectively, by fitting a Gaussian model on the planet. The opacity was measured and corrected for by using NIKA total power data as a tau-meter as described in \cite{catalano2014}. The Compton $y$ to surface brightness (measured in Jy/beam) conversions were computed by integrating the tSZ spectrum (see Sect.~\ref{sec:analysis}) within the NIKA bandpasses transmissions and accounting for the beam angular coverage, as described in more detail in \cite{adam2013}. For this campaign, the conversions are $-10.9 \pm 0.8$ and $3.5 \pm 0.5$ Jy/beam per unit of $y$ at 150 and 260~GHz, respectively, including the overall calibration error and the 2\% error arising from the bandpasses uncertainties.

%---------- FXD table
In Table~\ref{tab:instru}, we summarize the instrumental properties of the NIKA camera as it was used during \mbox{CL~J1226.9+332} observation.
\begin{table}
\caption{Instrumental characteristics of NIKA for the February 2014 campaign. See text for details.}
\begin{center}
\begin{tabular}{ccc}
\hline
\hline
Observing band & 150~GHz & 260~GHz \\
\hline
Gaussian beam model FWHM (arcsec) & 18.2 & 12.0 \\
Field-of-view (arcmin) & 1.9 & 1.8 \\
Effective number of detectors & 117 & 136 \\
Sensitivity (mJy/beam s$^{1/2}$)& 14 & 35 \\
Compton parameter to Jy/beam & -10.9 $\pm$ 0.8 & 3.5 $\pm$ 0.5 \\
Pointing errors (arcsec) & \textless 3 & \textless 3 \\
Calibration uncertainties & 7\% & 12\% \\
\hline
\end{tabular}
\end{center}
\label{tab:instru}
\end{table}

%========== Data reduction
\subsection{Data reduction}\label{sec:data_reduction}
%---------- Pipeline
The details of the NIKA data reduction are available in \cite{adam2013} and \cite{catalano2014}. Here, the main procedure is briefly summarized for the reader's convenience. Invalid detectors were removed based on the statistical properties of their noise and their optical response. Cosmic ray impacts on the arrays were flagged and removed from the data. To remove the low-frequency atmospheric emission from the data, a common-mode template was built by averaging the detector time stream across each array. This was done by flagging the source in signal-to-noise in an iterative manner to avoid ringing and reduce signal filtering effects. This data reduction was preferred for \mbox{CL~J1226.9+3332}, with respect to the spectral dual-band noise decorrelation described in~\cite{adam2013}. The latter allows more extended emission to be recovered but is noisier. Moreover, the cluster is sufficiently compact for any filtering effect to be weak enough, allowing the recovery of cluster maps at both wavelengths simultaneously. Frequency lines produced by the pulse tube of the cryostat were notch-filtered in the Fourier domain. Data were finally projected onto 2-arcsec pixel grid maps using inverse variance weighting and a nearest grid projection.

%---------- The map
The raw (direct output of the pipeline) 150 and 260~GHz NIKA maps of \mbox{CL~J1226.9+3332} are presented in Fig.~\ref{fig:CL1226map}. They are centered on the X-ray peak coordinates \citep[taken from][]{cavagnolo2009}, (R.A.,~Dec.)~=~(12h~26m~58.37s,~33$^o$~32'~47.4"), and have been smoothed with a 10 arcsec Gaussian filter for display purposes. The 150~GHz map shows a strong tSZ decrement reaching 18$\sigma$ per beam at the peak. Because the noise is higher at 260~GHz, and the tSZ signal weaker by a factor of about one third, the map does not show a very significant tSZ detection, even if diffuse positive emission is seen at about 3$\sigma$ on the map at the cluster position. However, the 260~GHz channel reveals the presence of a point source (referred to as PS260 in the following) located about 30 arcsec southeast of the \mbox{X-ray} center, detected at about 10$\sigma$. The source is not clearly detected at 150~GHz due to the strong tSZ signal, but it is visible as a lack of tSZ at its corresponding location. In Sect.~\ref{sec:point_source}, we discuss the implications of point source contamination on our tSZ observation.

%---------- The 2mm profile
In Fig.~\ref{fig:CL1226profile2mm}, we provide the flux density profile corresponding to the 150~GHz map. It is computed by averaging the signal in concentric annuli with the \mbox{X-ray} center taken as the origin. The profile appears to be smooth and peaks at the center.
	
%========== Transfer function
\subsection{Transfer function}\label{sec:transfer_function}
%---------- Transfer function in general
The data reduction described above induces an attenuation of the astrophysical signal in the recovered maps of Figure~\ref{fig:CL1226map}, since detectors are combined to remove the correlated noise. The characterization of this effective transfer function as a function of the angular scales was done by using noise plus input signal simulations. To do so, the map of an input known simulated astrophysical signal (see below) was compared to the output signal after processing.

%---------- The input signal
The simulated input signal was the one expected for clusters of galaxies, as described in detail in \citet{adam2013}. It was computed using a generalized Navarro, Frenk \& White (gNFW) pressure profile \citep[see Sec.~\ref{sec:icm_param_p},][]{nagai2007} integrated along the line of sight to produce a tSZ flux density map. The typical amplitude and angular size of the simulated clusters were similar to the one in Fig.~\ref{fig:CL1226map}, but the result transfer function was checked to see that it did not depend on the radial size and amplitude of the input signal.

%---------- The noise
To simulate the atmospheric and intrinsic correlated and uncorrelated noise, we used the NIKA data themselves. The real data used for the simulations were those of our other projects taken during the first NIKA open pool of February 2014, for which the scanning strategy was similar to that of \mbox{CL~J1226.9+3332}. These scans were taken with atmospheric conditions comparable to those during which the data presented in this paper were taken. The astrophysical signal within the data was checked to be faint enough that it did not affect the reduction, {\it i.e.} negligible compared to the noise. 

%---------- Residual
To deal with the residual noise contribution in the final processed maps, we considered the simulated data with and without including the known input signal. In this way we obtained estimates of both the processed noisy signal and of the noise itself, which we could subtract from the processed signal-plus-noise map to produce an output signal-only map. The transfer function was then computed as the ratio of the power spectra of the output signal, free of noise, and the input one. However, small noise residuals are observed because of the differences in the processing introduced by the signal itself.

\begin{figure}[h]
	\centering
	\includegraphics[width=0.45\textwidth]{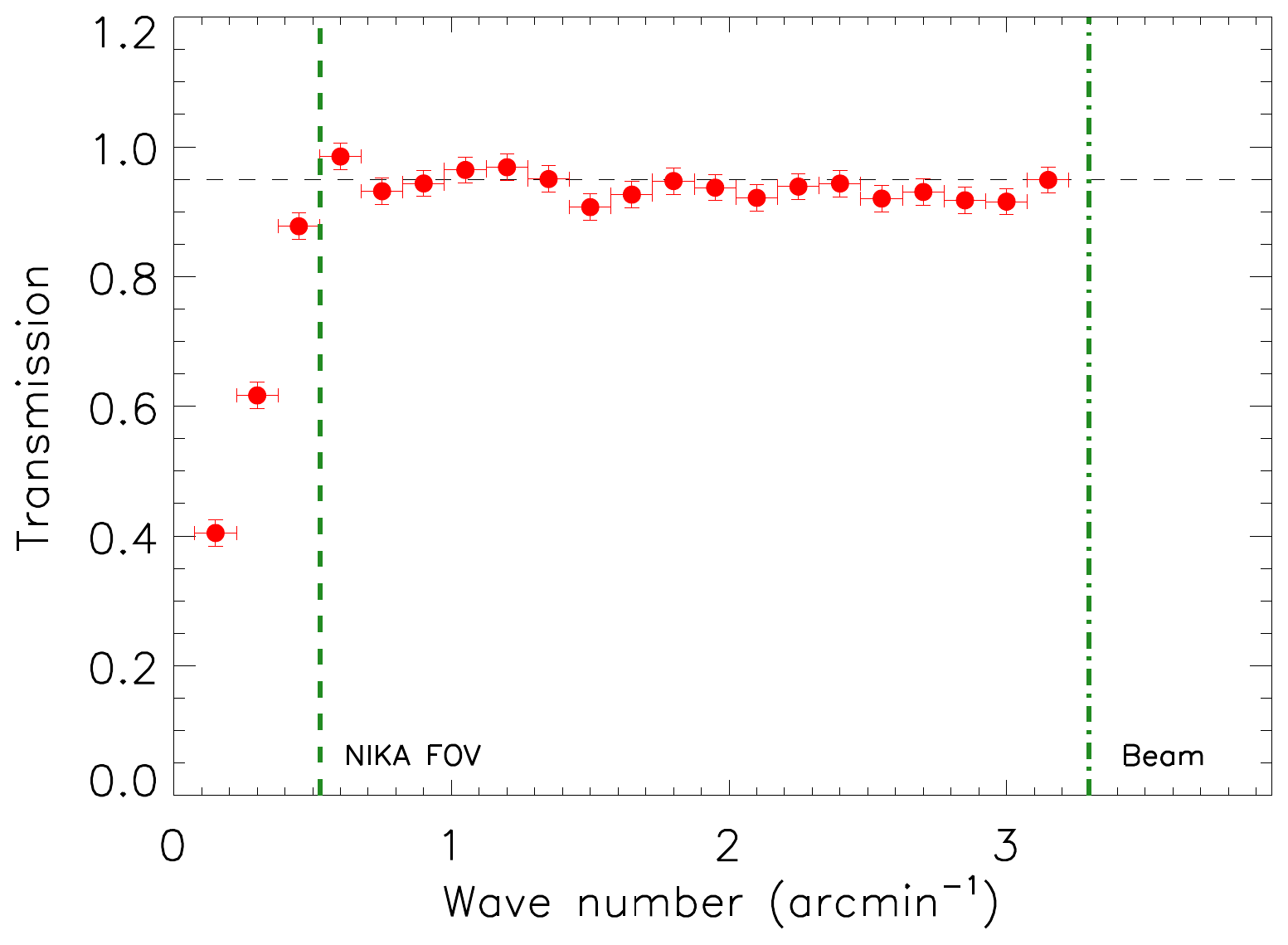}
	\caption{NIKA data reduction transfer function as a function of angular frequency. Uncertainties were computed using the dispersion of the results over the different noise realizations. The 150~GHz beam cutoff and size of the NIKA field of view are also represented by green dashed lines for illustration, {\it i.e.} (18.2 arcsec)$^{-1}$ and (1.9 arcmin)$^{-1}$. The black horizontal dashed line corresponds to 5\% filtering.}
        \label{fig:transfer_function}
	\end{figure}

%---------- The obtained transfer function and its use
The estimated transfer function is given in Fig.~\ref{fig:transfer_function}. The uncertainties were calculated from the dispersion of the transfer function obtained for all the scans used to compute it, and are mostly due to residual noise. As we can see, it is approximately flat and close to one, with $\sim 5$ \% attenuation, on scales smaller than the NIKA field of view. On larger scales, the recovered flux vanishes smoothly with decreasing wave number. In Sec.~\ref{sec:param_estim}, we use this transfer function when comparing a model to the NIKA map.

%========== Point source contamination
\subsection{Point source contamination}\label{sec:point_source}
The SZA data were used to search for radio sources around \mbox{CL~J1226.9+3332} \citep{muchovej2007}. From their observations, no such objects are present within the NIKA field. From the residual between the MUSTANG map and the SZA pressure model of \cite{mroczkowski2009}, \cite{korngut2011} have inferred the presence of a possible submillimeter source 10 arcsec north of the \mbox{X-ray} peak. Using the NIKA 260~GHz frequency band, we searched for such a contaminant and do not observe any point source within a 1.5 arcmin radius around the map center, apart from PS260. This is done by fitting PS260 and the tSZ signal simultaneously, as described in detail in Sect.~\ref{sec:param_estim}. The root-mean-squared between the best-fit model and the data allows us to set a 2$\sigma$ flux upper limit of 1.5~mJy at this frequency. Therefore, the feature seen by \cite{korngut2011} is unlikely to be a real submillimeter source. The flux distribution of the detected point source, PS260, is fitted using a Gaussian model with FWHM fixed to the NIKA 260~GHz beam. Its flux is measured to be 6.8 $\pm 0.7$~(stat.)~$\pm 1.0$~(cal.)~mJy and its position (R.A.,~Dec.)~=~(12h~27m~0.01s,~33$^o$~32'~42.0"), with statistical and calibration error quoted as stat and cal. We note that the source possibly coincides with two known optically detected galaxies J12265995+3332405 and J12265923+3332405~\citep{holden2009}. They are located 1.7 and 9.8 arcsec away from the best-fit position obtained, respectively. We conclude that no additional point sources, wether radio or submillimeter, affect the 150~GHz map of NIKA. The detected source is accounted for in the analysis described hereafter.

%###############################################################################################
%##########################                                  ANALYSIS                                   ##########################%###############################################################################################
\section{Characterization of the intracluster medium}\label{sec:analysis}
%========== tSZ and Xray observables
\subsection{SZ and X-ray observables}\label{sec:observables}
%---------- tSZ give P_e
\subsubsection{Thermal SZ}\label{sec:tsz}
The tSZ effect results in a distortion of the CMB black-body spectrum whose frequency dependence is given by \citep{birkinshaw1999}
\begin{equation}
	f(x, T_e) = \frac{x^4 e^x}{\left(e^x-1\right)^2} \left(x  \ \mathrm{coth}\left(\frac{x}{2}\right) - 4\right) \left( 1 + \delta_{tSZ}(x, T_e) \right), 
	\label{eq:sz_f_x}
\end{equation}
where $x = \frac{h \nu}{k_{\mathrm{B}} T_{\mathrm{CMB}}}$ is the dimensionless frequency, $h$ the Planck constant, $k_{\mathrm{B}}$ the Boltzmann constant, $\nu$ the observation frequency, and $T_{\mathrm{CMB}}$ the temperature of the CMB. We use the \cite{itoh1998} relativistic corrections to compute $\delta_{tSZ}(x, T_e)$, where $T_e$ is the electrons temperature. The induced change in intensity relative to the primary CMB intensity $I_0$ reads as
\begin{equation}
	\frac{\Delta I_{tSZ}}{I_0} = y \ f(x, T_e) \ ,
\label{eq:deltaI}
\end{equation}
 where $y$ is the Compton parameter, which measures the integrated electronic pressure $P_{\mathrm{e}}$ along the line of sight, $dl$, written as
   \begin{equation}
	y = \frac{\sigma_{\mathrm{T}}}{m_{\mathrm{e}} c^2} \int P_{\mathrm{e}} dl.
	\label{eq:y_compton}
   \end{equation}
The parameter $\sigma_{\mathrm{T}}$ is the Thomson cross section, $m_{\mathrm{e}}$ is the electron rest mass, and $c$ the speed of light. Neglecting the relativistic corrections, the tSZ spectral distortion is nil around 217~GHz, negative below and positive above.

%---------- Velocity
\subsubsection{Kinetic SZ}\label{sec:ksz}
In addition to the tSZ effect, the kinetic Sunyaev-Zel'dovich (kSZ) is caused by the motion of the intracluster gas and its electrons relative to the CMB. This motion leads to a Doppler shift of the CMB photons that are scattered via the Compton effect. It can be expressed as \citep{birkinshaw1999}
\begin{equation}
\frac{\Delta I_{kSZ}}{I_0} = g(x, v_z, T_e) \ \sigma_T \frac{-v_z}{c} \int n_e dl,
	\label{eq:ksz}
\end{equation}
where $v_z$ is the line-of-sight peculiar velocity of the cluster with respect to the Hubble flow, which is positive (negative) for a cluster receding from (coming towards) the observer, and $n_e$ the electronic density. The function $g(x, v_z, T_e)$ provides the spectral dependence of the kSZ effect as 
\begin{equation}
	g(x, v_z, T_e) = \frac{x^4 e^x}{\left(e^x-1\right)^2} \left( 1 + \delta_{kSZ}(x, v_z, T_e) \right).
	\label{eq:ksz_spec}
\end{equation}
Again, we use \cite{itoh1998} to account for the relativistic corrections $\delta_{kSZ}$ and neglect the velocity dependance as it is expected to be less than 1\%. By writing $\Delta I (\nu) = A_{tSZ} \Delta I_{tSZ}(\nu) + A_{kSZ} \Delta I_{kSZ}(\nu)$ where $A_{tSZ,kSZ}$ stand for dimensionless amplitudes (see Eqs.~\ref{eq:deltaI} and \ref{eq:ksz}), and assuming the observed region to be isothermal, one can deduce the line-of-sight velocity as $v_z = -\frac{A_{kSZ} \ k_B T_e}{A_{tSZ} \ m_e c}$.

%---------- X-ray gives n_e
\subsubsection{X-ray emission}\label{sec:xray}
The \mbox{X-ray} surface brightness, in units of counts cm$^{-2}$ s$^{-1}$ sr$^{-1}$, is related to the electronic density as
\begin{equation}
	S_X = \frac{1}{4 \pi (1+z)^4} \int n_e^2 \Lambda(T_e, Z) dl.
\label{eq:SX}
\end{equation}
The parameter $z$ is the redshift, $\Lambda(T_e, Z)$ is the cooling function that is proportional to $T_e^{1/2}$, and $Z$ is the metallicity. Additionally, the gas temperature can be estimated from \mbox{X-ray} spectroscopy. 

%========== Parametrization of the ICM
\subsection{Intracluster medium modeling}\label{sec:icm_parametrization}
%---------- Pressure
\subsubsection{Pressure profile}\label{sec:icm_param_p}
%The model
The cluster electronic pressure distribution is modeled by a spherical gNFW profile \citep{nagai2007}, described by
\begin{equation}
	P_e(r) = \frac{P_0}{\left(\frac{r}{r_p}\right)^c \left(1+\left(\frac{r}{r_p}\right)^a\right)^{\frac{b-c}{a}}}.
\label{eq:gNFW}
\end{equation}
The parameter $P_0$ is a normalizing constant; $r_p$ is a characteristic radius; and $a$, $b$, and $c$ set the slopes at intermediate, large, and small radii, respectively. We can also write $P_0 = P_{\Delta} \times \mathds{P}_0$ and $r_{\Delta} = c_{\Delta} \ r_p$, where $P_{\Delta}$ is the average pressure within $r_{\Delta}$, $\mathds{P}_0$ is a normalizing constant, and $c_{\Delta}$ the concentration parameter \citep{arnaud2010}. The mass enclosed within $r_{\Delta}$, $M(r = r_{\Delta})$, is then related to $P_{\Delta}$ by a scaling law. One can finally define $\theta_{p,\Delta} = r_{p,\Delta} / D_A$, where $D_A$ is the angular distance of the cluster.

%What we do with the pressure parameters
In the following, we use three different choices to fix the slope parameters (see Table~\ref{tab:table_pressure_models}). 1) We fix $c$ to the value obtained by \cite{comis2011} for this cluster and fix $a$ and $b$ to the one obtained by \cite{planck2013pressure_profile} when stacking the tSZ signal of 62 nearby clusters. This choice is used as the baseline since the two outer slope parameters have been obtained directly from tSZ data and are expected to provide a good description of most clusters. The parameter $c$ was not fitted by \cite{planck2013pressure_profile} so we rely on Chandra \mbox{X-ray} data that are specific to \mbox{CL~J1226.9+3332}. This set of parameters is referred to PPC in the following. 2) We fix $a$, $b$, and $c$ to the values obtained by \cite{nagai2007b} based on \mbox{X-ray} Chandra clusters and numerical simulations. This set of parameters allows us to directly compare our results to that of \cite{mroczkowski2009}, who used them in their modeling. It is referred to as NNN. 3) We fix $b$ and $c$ to values similar to those of PPC, but fit for the parameter $a$ since it corresponds to scales at which NIKA is the most sensitive for this cluster. This choice is referred to as FPC. 

%Discussion
The parameters $P_0$ and $r_p$ are always allowed to vary. Since the pressure profile parameters are highly degenerate, in particular $a$, $P_0$, and $r_p$, the main difference between the three models relies on the choice of the core slope parameter $c$ and the outer slope parameter $b$, which cannot be constrained directly with the NIKA data. As discussed in Sect.~\ref{sec:results}, the core slope parameter is related to the thermodynamics of the cluster core but does not affect the overall mass determination. By contrast, the outer slope is associated to the steepness of the mass profile at outer radii. Therefore it controls the cluster overall mass at large radii and can lead to a bias in the mass estimate for radii above those constrained directly from the data. By using these three models, we test the impact of the choice of the model in the regions of the ICM profiles for which NIKA is not directly sensitive.
\begin{table}
\caption{Pressure profile parameters for the three models presented in this paper.}
\begin{center}
\begin{tabular}{cccc}
\hline
\hline
Model label & $a$ & $b$ & $c$ \\
\hline
PPC & 1.33 & 4.13 & 0.014 \\
NNN & 0.9 & 5.0 & 0.4 \\
FPC & free & 4.13 & 0.014\\
\hline
\end{tabular}
\end{center}
\label{tab:table_pressure_models}
\end{table}

%---------- Density
\subsubsection{Density profile}\label{sec:icm_param_n}
Following \cite{mroczkowski2009} and since we use the work of \cite{comis2011}, the electron density profile is described by a simplified version (SVM) of the model suggested by \cite{vikhlinin2006}
\begin{equation}
	n_e(r) = n_{e0} \left[1+\left(\frac{r}{r_c}\right)^2 \right]^{-3 \beta /2} \left[ 1+\left(\frac{r}{r_s}\right)^{\gamma} \right]^{-\epsilon/2 \gamma},
\label{eq:SVM}
\end{equation}
which is an extension of the $\beta$ model \citep{cavaliere1978} with an additional steepening freedom at radii larger than $\sim r_s$, for which the slope parameter is $\epsilon$. The core radius is still given by $r_c$, and $\gamma$ accounts for the width of the transition between the two profiles. In the case of $\epsilon = 0$, this model is equivalent to the standard $\beta$ model. Similar to \cite{mroczkowski2009} and \cite{comis2011}, we fix the parameter $\gamma = 3$ because it is a good fit to all the clusters considered by \cite{vikhlinin2006}, and leave the other ones as free parameters.

%---------- Temperature and entropy
\subsubsection{Temperature and entropy}\label{sec:icm_param_tk}
Assuming the ideal gas law, the temperature of the electron population can simply be computed as
\begin{equation}
	k_B \ T_e(r) = P_e(r) / n_e(r).
	\label{eq:temp_profile}
\end{equation}
It is implicitly modeled as the ratio of the distributions given by the gNFW and SVM models of Eqs.~\ref{eq:gNFW} and ~\ref{eq:SVM}.
The ICM entropy is defined as \citep[see review from][]{voit2005}
\begin{equation}
	K(r) =  \frac{P_e(r)}{n_e(r)^{5/3}}.
	\label{eq:entropy_profile1}
\end{equation}

%---------- Mass
\subsubsection{Mass distribution}\label{sec:icm_param_mass}
When assuming \mbox{CL~J1227.9+3332} to be in hydrostatic equilibrium, its total mass enclosed within $r$, $M_{tot}(r)$, is related to the electronic density and pressure profiles through
\begin{equation}
	\frac{dP_e(r)}{dr} = -\frac{\mu_{gas} m_p n_e(r) G M_{tot}(r)}{r^2},
	\label{eq:hse}
\end{equation}
where $m_p$ is the proton mass and $G$ the Newton's constant. We assume in this paper a mean molecular weight $\mu_e = 1.15$ for the electrons and $\mu_{gas} = 0.61$ for the gas. By directly integrating the electronic density profile up to a radius $R$, we obtain the gas mass enclosed within $R$,
\begin{equation}
	M_{gas}(R) = 4 \pi \int_0^R \mu_e m_p n_e(r) r^2 dr.
	\label{eq:gas_mass}
\end{equation}
It is straightforward to deduce the gas fraction profile, defined as the ratio at a given radius between the gas mass and the total mass enclosed within $r$, as
\begin{equation}
	M_{gas}(r) = f_{gas}(r) M_{tot}(r).
	\label{eq:gas_frac}
\end{equation}
Finally, the total mass is directly related to $R_{\Delta}$ from its definition as $M_{tot}(r_{\Delta}) = \frac{4}{3} \pi \rho_c(z) \Delta r_{\Delta}^3$, where $\rho_c(z)$ is the critical density of the Universe at redshift $z$. Combining the value of $R_{\Delta}$ and $r_p$, directly related to the pressure profile, we can therefore measure the concentration parameter $c_{\Delta}$.

%========== Extra data
\subsection{Extra datasets}\label{sec:extra_data}
In addition to the NIKA data we consider the ACCEPT and Planck data sets.
%---------- Density from ACCEPT
\subsubsection{ACCEPT density profile}\label{sec:accept}
We make use of the ACCEPT catalog \citep[Archive of Chandra Cluster Entropy Profile Tables\footnote{\url{http://www.pa.msu.edu/astro/MC2/accept/}},][]{cavagnolo2009}. We only consider the deprojected \mbox{X-ray} density profile of \mbox{CL~J1226.9+3332}, which is computed from the publicly available Chandra data. The angular resolution of Chandra, $\sim 0.5$ arcsec, is negligible compared to that of NIKA. As fully explained in \cite{cavagnolo2009}, the flux measured in the energy range 0.7--2.0 keV is a good diagnosis of the ICM density (Eq.~\ref{eq:SX}). The high angular resolution surface brightness profile is therefore converted into a deprojected electron density profile using normalization and count rates taken from the spectral analysis. The profile extends up to 835~kpc, which corresponds approximately to $R_{500}$ (see Sect.~\ref{sec:results}).

%---------- Y_tot from Planck
\subsubsection{Planck integrated Compton parameter}\label{sec:planck}
The cluster \mbox{CL~J1226.9+3332} is not in the Planck tSZ cluster catalog \citep{planck2013catalogue} since its flux is diluted by the Planck beam, and it is therefore not detected with high enough signal-to-noise. Nevertheless, in addition to NIKA tSZ observations, we use the Planck maps to produce a Compton $y$ parameter map as described in \cite{planck2013ymap} (see \cite{hurier2013} for the method). Its angular resolution is 7.5 arcmin, limited by the lowest Planck frequency channel used to construct it. This map is used to measure the integrated Compton parameter of \mbox{CL~J1226.9+3332} within $\theta_{\rm max}$, defined as
\begin{equation}
	Y_{\theta_{\rm max}} = \int_{\Omega(\theta_{\rm max})} y \ d\Omega.
\label{eq:Yint}
\end{equation}
The uncertainty on this quantity is obtained by applying the same integration on the map at positions around the cluster, where the noise is homogeneous and the map is free of emission. We also check on the jackknife \citep[half-ring half-difference, see][]{planck2013mission} map that the error is consistent with the expected noise. We obtain $Y_{\Omega(15^{'})}~=~\left(0.94~\pm~0.36\right)~\times~10^{-3}$ arcmin$^2$.

%========== MCMC
\subsection{Maximum likelihood analysis}\label{sec:param_estim}
%---------- MCMC sampling
We aim at recovering the three-dimensional electronic pressure and electronic density profiles of \mbox{CL~J1226.9+3332}. To do so, we use an approach in which input models are processed similarly to the measured tSZ signal such that they can be compared to it. Best-fit values of the electronic pressure and density model parameters are jointly obtained from a Markov Chain Monte Carlo (MCMC) approach, using a Metropolis-Hasting algorithm \citep{chib1995}. A set of chains of tested models samples the multidimensional likelihood parameter space. At each step of the chains, a model map is computed by integrating the tested pressure model along the line of sight. The map is then convolved with the NIKA beam and the pipeline transfer function. The model is converted into surface brightness using the Jy/beam to $y$ conversion factors given in Sect~\ref{sec:data_reduction}. A radial temperature model is inferred from the pressure and the density models, and used to account for relativistic corrections~\citep{itoh1998} on the tSZ map model. We also use the Planck integrated flux to add an extra constraint on the overall flux. The flux and the position of PS260 are simultaneously fitted. We only impose a Gaussian prior on its position based on the fit at 260~GHz, assuming a 3 arcsec uncertainty. We include a set of nuisance parameters in the fit, such as calibration uncertainties, map zero level, and the pointing position, which are randomly sampled within their error bars at each step. 

%---------- chi2
The $\chi^2$ used in the Metropolis-Hasting acceptance (or rejection) process of the chain samples is defined as
\begin{multline}
\chi^2 = \chi^2_{NIKA} + \chi^2_{ACCEPT} + \chi^2_{Planck} \\
          = \sum_{i=1}^{N_{pix}} \left(\frac{M^{NIKA}_i - M^{model}_i}{\sigma_i^{NIKA}}\right)^2 \\
          + \sum_{j=1}^{N_{bin}} \left(\frac{n_e(r_j)^{ACCEPT} - n_e(r_j)^{model}}{\sigma_j^{ACCEPT}}\right)^2 \\
          + \left(\frac{Y^{Planck}_{\theta_{\rm max}} - Y^{model}_{\theta_{\rm max}}}{\sigma^{Planck}}\right)^2
\label{eq:chi2}
\end{multline}
where the sums are made over the number of pixels in the NIKA map ($N_{pix}$) and the number of radial bins of the ACCEPT density profile ($N_{bin}$). The quantity $M$ represents the 150~GHz only tSZ surface-brightness-plus-point-source map. The parameters $\sigma^{NIKA}$, $\sigma^{ACCEPT}$ and $\sigma^{Planck}$ are the respective errors, assumed to be Gaussian. While \mbox{X-ray} counts follow Poisson statistics, the deprojected density profile is computed by combining a set of random variables; therefore, assuming that the central limit theorem applies in this process, we expect the error statistics followed by the ACCEPT deprojected density profile to be Gaussian. It is also naturally the case for Planck and NIKA errors. 

%---------- Convergence and distribution
The convergence of the MCMC is ensured by the \cite{gelman1992} test. Once reached, the histogram of the chains along the considered parameter is marginalized over all other dimensions (including nuisance parameters) providing the posterior probability distribution for each fitted parameter. The integrated posterior probability distribution up to 68\% probability gives the quoted errors.

%---------- Error
To estimate the uncertainties on the derived cluster physical properties for each radial bin, we fully propagate the information contained in the MCMC parameter chains to the given quantity. For each set of parameters tested against the data ({\it i.e.}, a model), we compute all derived physical quantities as a function of the radial distance. Therefore, for each radius, we obtain a probability distribution function for the considered quantity. We then compute the reference value of the given quantity as the median of the distribution and its error by integrating the distributions up to the requested confidence limit, for each radial bin.

%###############################################################################################
%##########################                                   RESULTS                                   ##########################%###############################################################################################
\section{Results and discussions}\label{sec:results}
%========== Morphology
\subsection{NIKA dual-band detection and mapping of the tSZ signal}\label{sec:dual_band_detection}
%---------- y profile and 1mm detection
	\begin{figure}[h]
	\centering
	\includegraphics[width=0.45\textwidth]{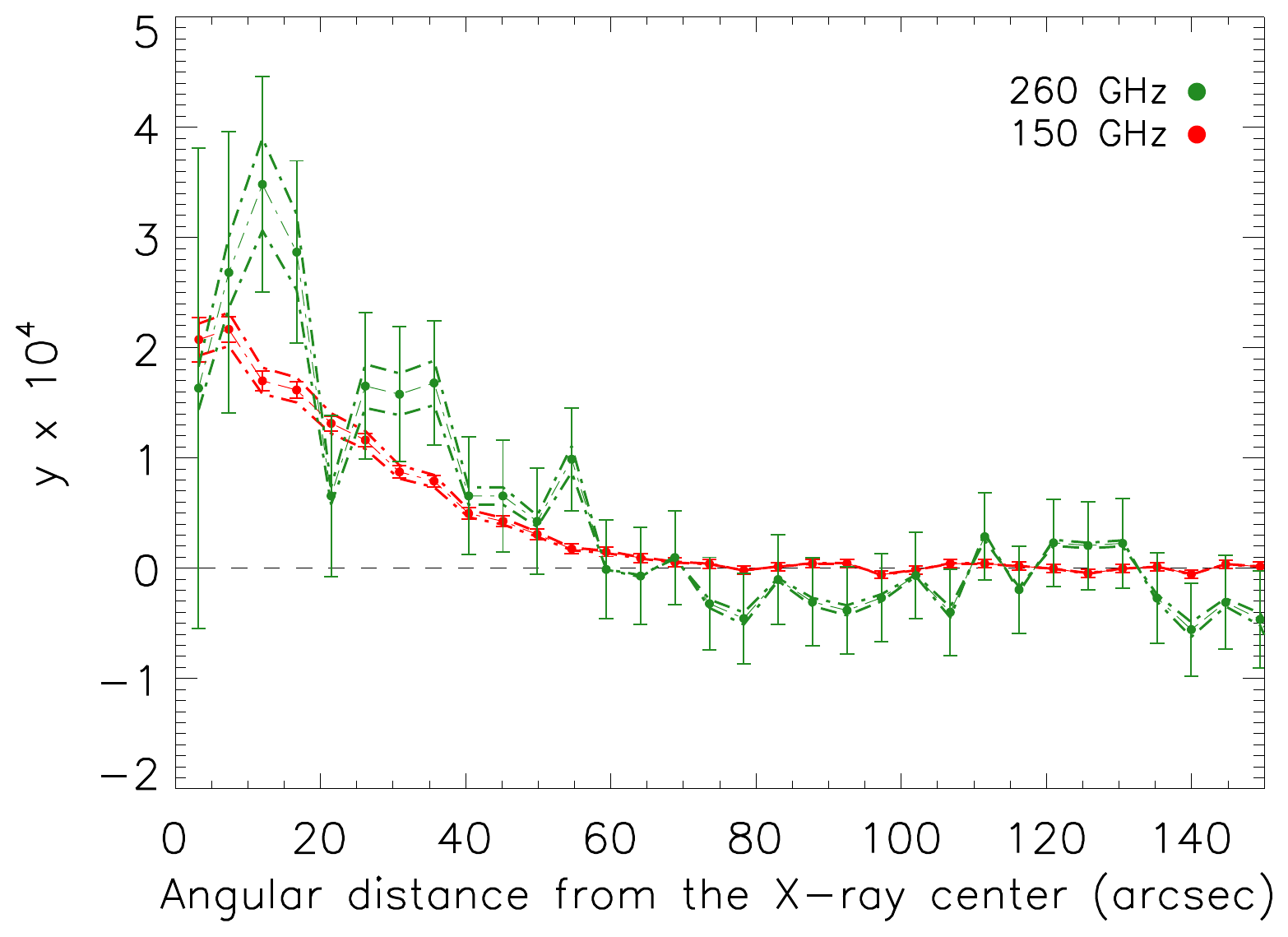}
	\caption{Compton parameter profile $y$ of \mbox{CL~J1226.9+3332} at 150~GHz (red) and 260~GHz (green). The point source has been subtracted before extracting the profile. Statistical uncertainties are shown as error bars, and systematic uncertainties are given as a dashed-line envelope.}
        \label{fig:y_profile}
	\end{figure}
From the MCMC analysis we obtain a flux of $1.9\pm 0.2$ (stat.) mJy at 150~GHz for PS260. With these results, PS260 is subtracted from the maps in the following analysis. The Compton parameter profile is computed by averaging the signal within radial bins and accounting for the conversion between flux density and Compton parameter. Figure~\ref{fig:y_profile} shows the Compton parameter radial profile, computed from the \mbox{X-ray} center for both 150 and 260~GHz. The signal is detected on the profile up to about 1 arcmin at 150~GHz. Error bars are only statistical uncertainties; calibration uncertainties would result in an overall multiplicative factor to apply to the entire profile. The two profiles are compatible over the whole radial range. Nevertheless, we notice the presence of a few distinct peaks in the 260~GHz profile. These peaks might indicate the presence of additional contamination from submillimeter sources, which are most probably below the noise in the 260~GHz map. Assuming a dust-like spectrum, we expect the contribution from these sources to be negligible at 150~GHz. By fitting the 260~GHz profile to the 150~GHz one, taken as the model, we obtain a 7$\sigma$ tSZ detection at 260~GHz. We obtain a reduced $\chi^2$ of 0.95 with 24 degrees of freedom.

%---------- Residual maps
	\begin{figure*}[h]
	\centering
	\includegraphics[width=0.48\textwidth]{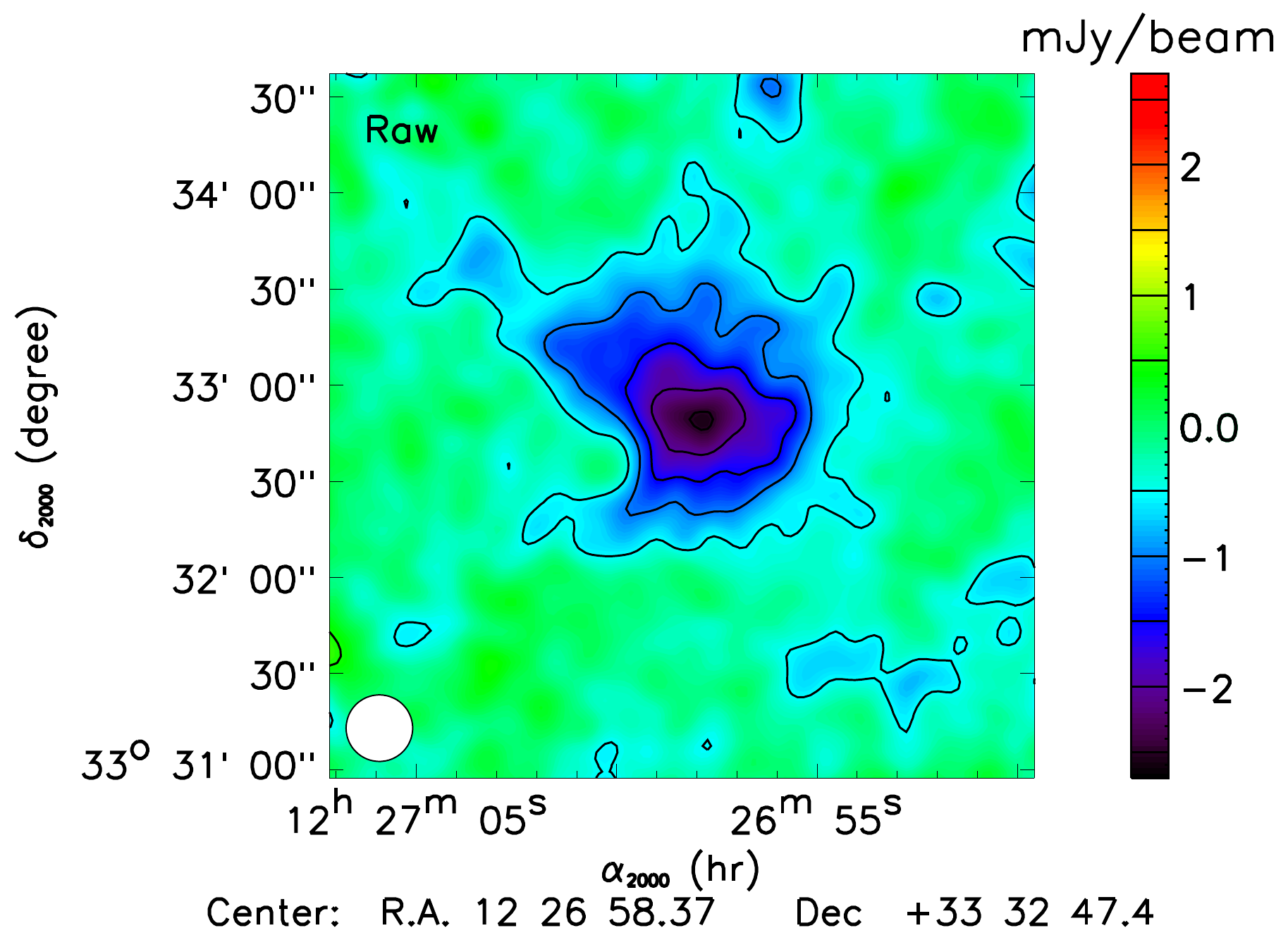}
	\includegraphics[width=0.48\textwidth]{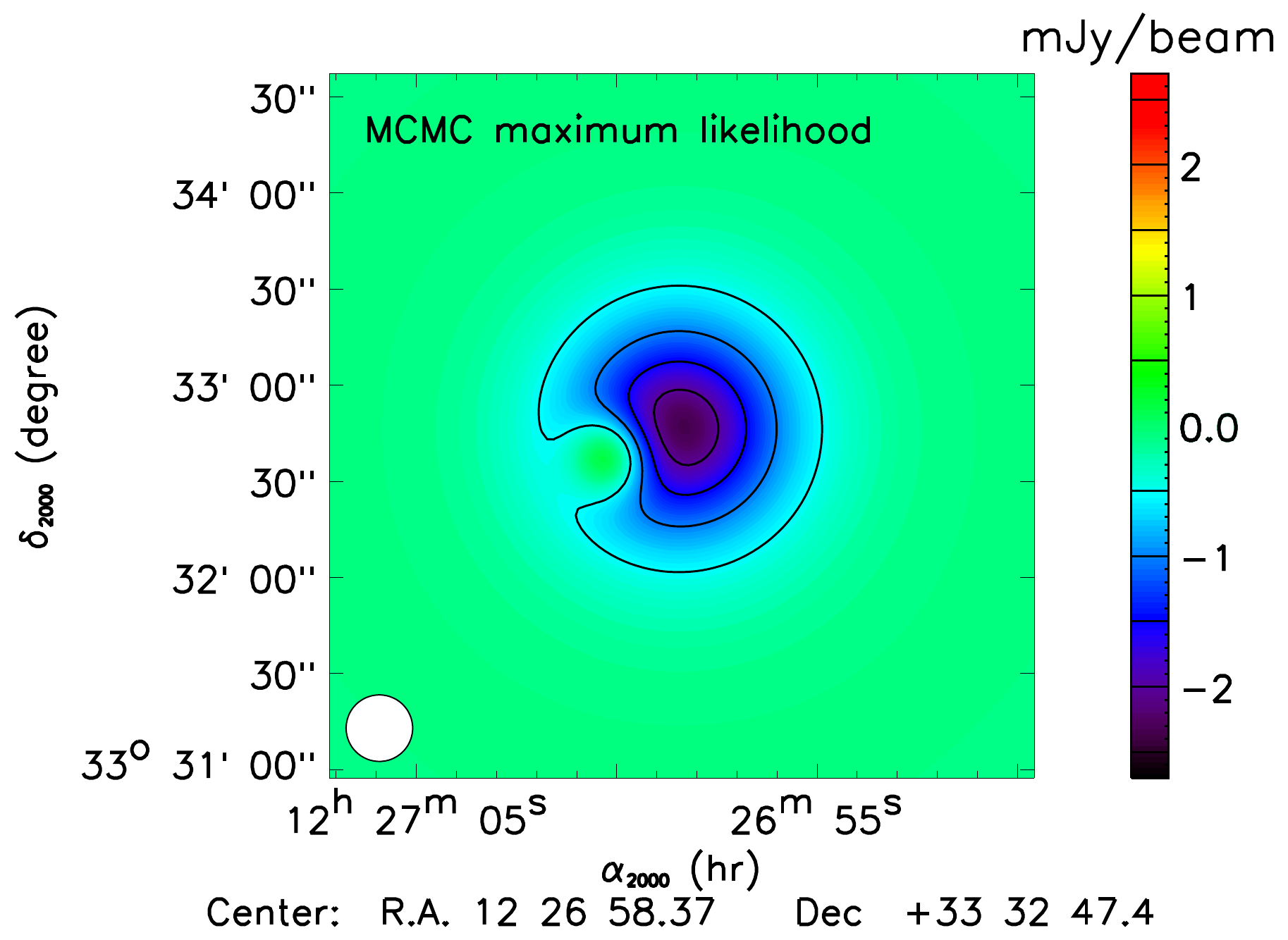}
	\includegraphics[width=0.48\textwidth]{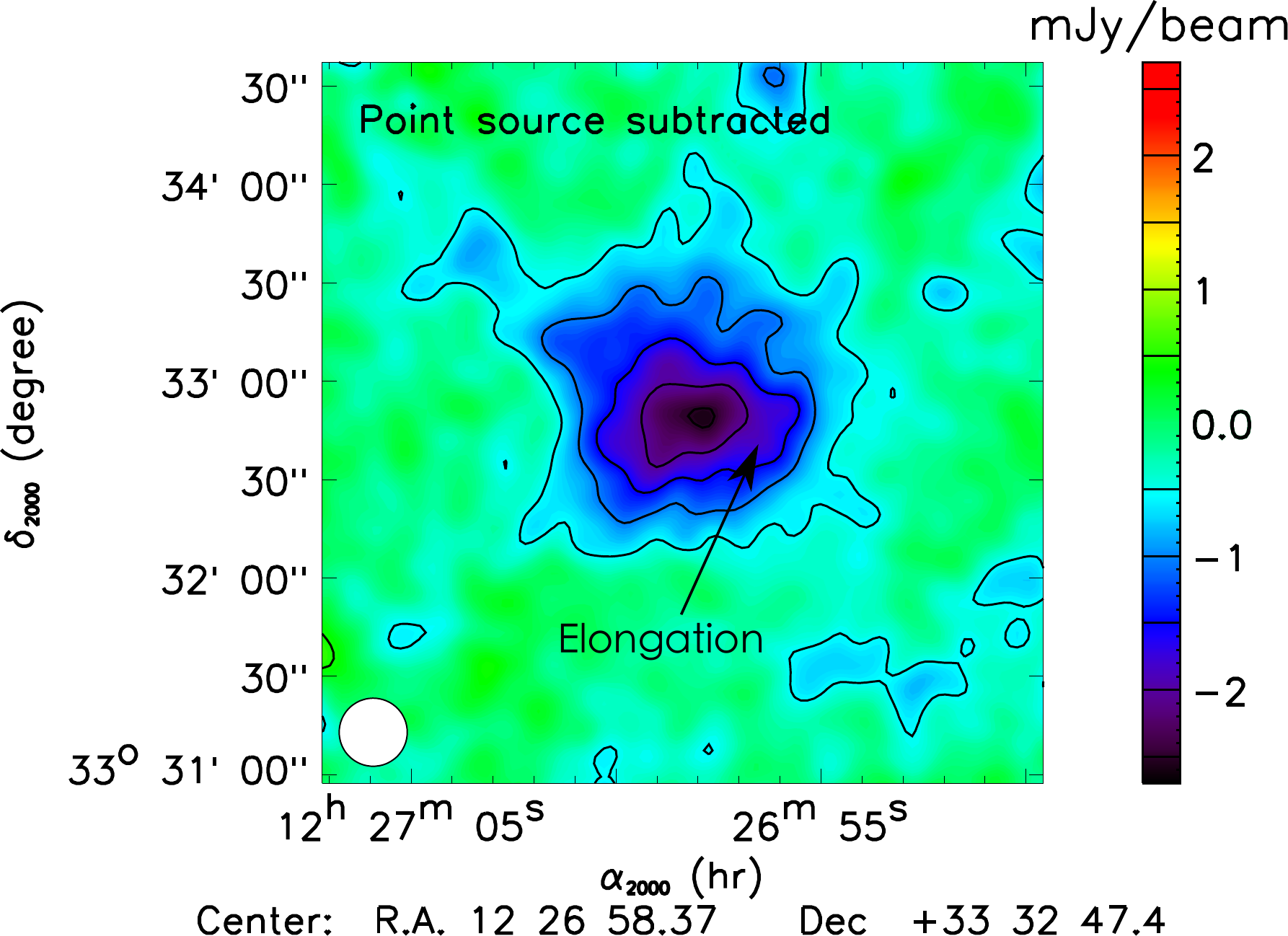}
	\includegraphics[width=0.48\textwidth]{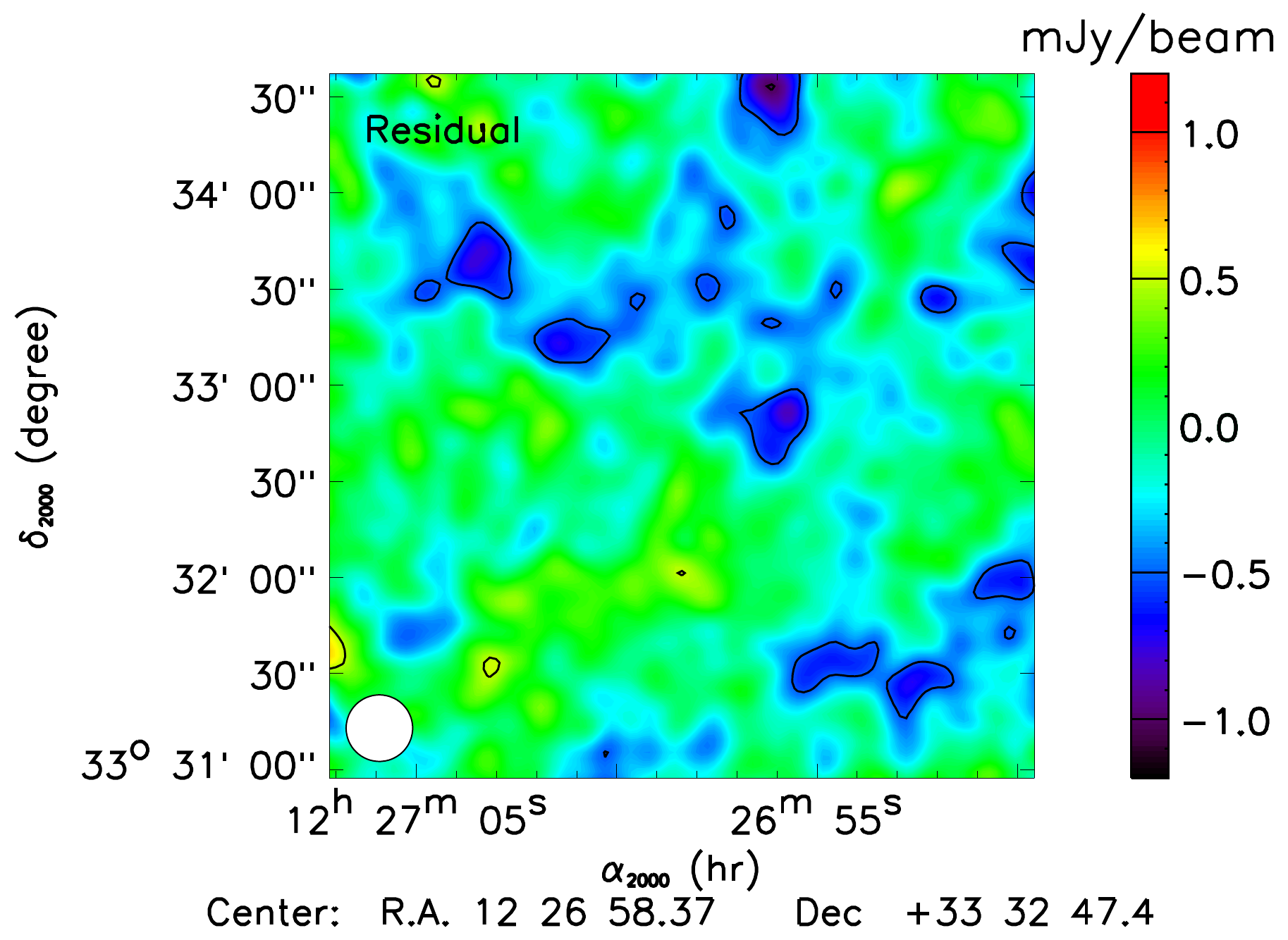}
	\caption{Top left: NIKA 150~GHz raw map of \mbox{CL~J1226.9+3332}. Top right: MCMC maximum likelihood tSZ + point source model. Bottom left: point-source-subtracted map. Bottom right: tSZ + point-source-subtracted residual map. The contours are spaced by 0.5~mJy/beam, and the maps have been smoothed with a 10 arcsec Gaussian filter. The effective beam is shown in the bottom left corner of each map.}
        \label{fig:best_fit_map}
	\end{figure*}
Figure~\ref{fig:best_fit_map} provides the raw, best-fit, point source subtracted and residual maps obtained from the maximum likelihood analysis. After subtracting PS260 the cluster appears circular at the NIKA resolution and is aligned with the \mbox{X-ray} peak on which the maps are centered. The signal is extended and clearly detected at the map level up to 1 arcmin scales. The NIKA map is morphologically consistent with previous interferometric observations by SZA \citep{joy2001,muchovej2007,mroczkowski2009} and does not show any evidence of being disturbed on large angular scales. However, the cluster core is slightly elongated toward the southwest at scales close to our our beam (and smaller). Using MUSTANG 90~GHz observations, at an effective resolution of 11 arcsec, \cite{korngut2011} have indeed detected a narrow ridge $\sim 20$ arcsec long located about 10 arcsec from the \mbox{X-ray} center towards the southwest. This is consistent with the hotter region found by \cite{maughan2007} using Chandra and XMM \mbox{X-ray} data. Additional lensing observations from HST \citep{jee2009} also reveals the presence of a secondary peak in the surface mass distribution in this region. 

The NIKA observations agree with \mbox{CL~J1226.9+3332} being relaxed on large scales with a disturbed core, the origin of the latter being probably due to the merger of a smaller subcluster. Since NIKA probes scales between $\sim 20$ arcsec to a few arcmin, these observations complement the ones by MUSTANG on small scales ($\sim$ 10 -- 50 arcsec) and by SZA interferometric data that are the most sensitive on scales of a few arcmin. Finally, we notice that the NIKA residual map is well correlated with the temperature map presented in \cite{maughan2007}, {\it i.e.} the tSZ signal appears to be slightly stronger on the north (being under estimated by our spherically symmetric model), where the gas is hotter than it is in the south (respectively, overestimated), where the gas is cooler.

%---------- kSZ
	\begin{figure}[h]
	\centering
	\includegraphics[width=0.48\textwidth]{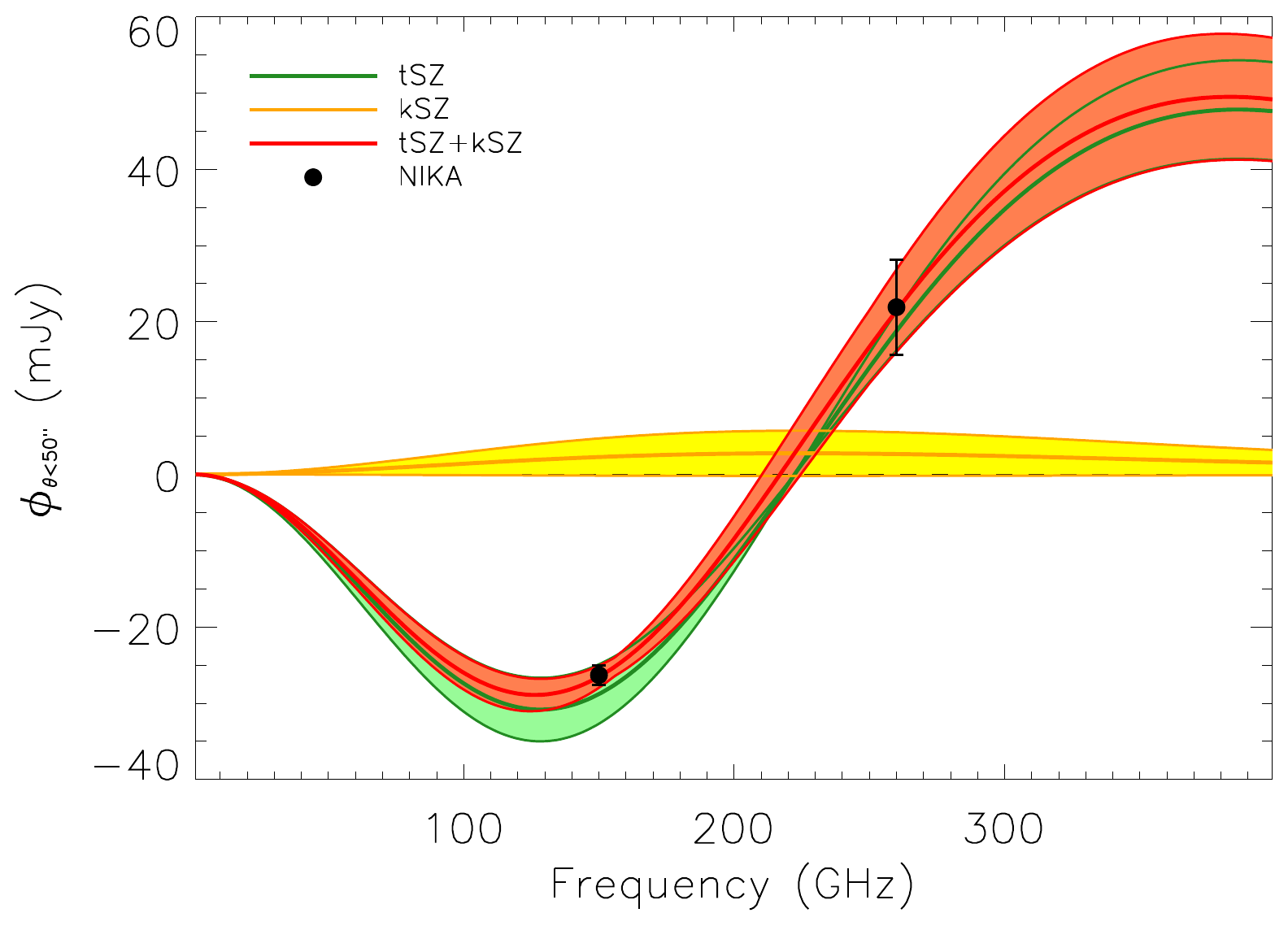}
	\caption{Constraints on the kSZ and tSZ spectra computed within 50 arcsec of the \mbox{X-ray} center. The green, yellow, and red swaths give the tSZ, kSZ, and tSZ+kSZ contributions, respectively. The two data points are the NIKA measurements.}
        \label{fig:ksz_spec}
	\end{figure}
By measuring the integrated flux toward the cluster within a 50 arcsec radius circle centered on the \mbox{X-ray} peak at both NIKA wavelengths, it is possible to set constraints on the kSZ contribution (see Sec.~\ref{sec:ksz}). As shown in Fig.~\ref{fig:ksz_spec}, the kSZ spectrum amplitude is compatible with zero within 1$\sigma$. Assuming the cluster average temperature within the region considered to be $T_e = 10 \pm 1$ keV, we infer a limit on the line-of-sight velocity of $v_z = -445 \pm 461$ km/s at 1$\sigma$ including calibration uncertainties.

%========== ICM thermo
\subsection{Intracluster medium radial distribution}\label{sec:icm_3d_result}
	\begin{figure}[h]
	\centering
	\includegraphics[width=0.48\textwidth]{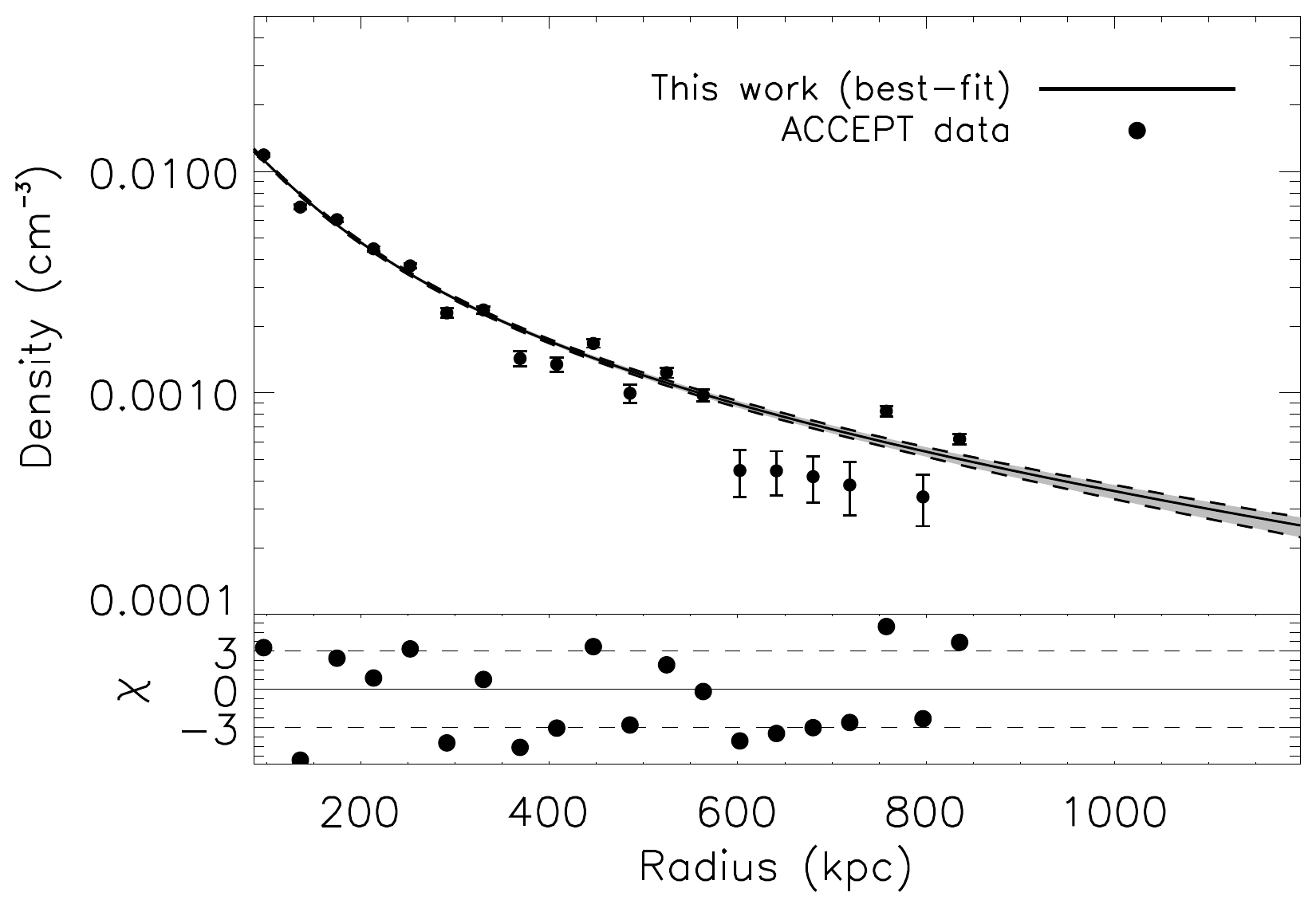}
	\caption{Density profile as a function of physical distance from the cluster center. The data correspond to those from the ACCEPT database \citep{cavagnolo2009} discussed in the text. The solid line represents the best-fit density model. The 1$\sigma$ uncertainties are represented by the gray contours. The difference between the data and the best-fit model normalized by the 1$\sigma$ uncertainties ($\chi$) is also shown.}
        \label{fig:density_profile}
	\end{figure}
	
	\begin{figure*}[h]
	\centering
	\includegraphics[height=4.3cm]{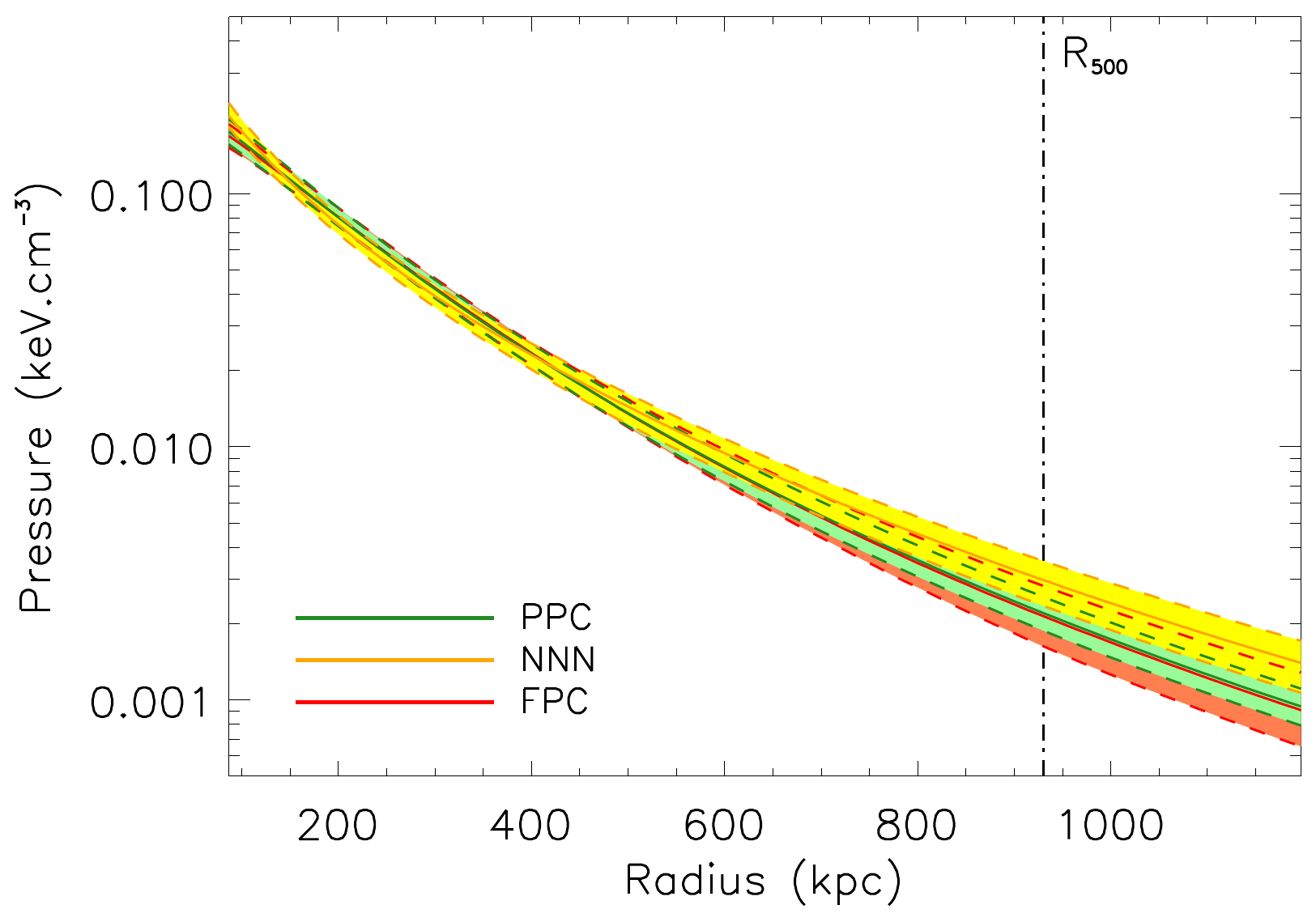}
	\includegraphics[height=4.3cm]{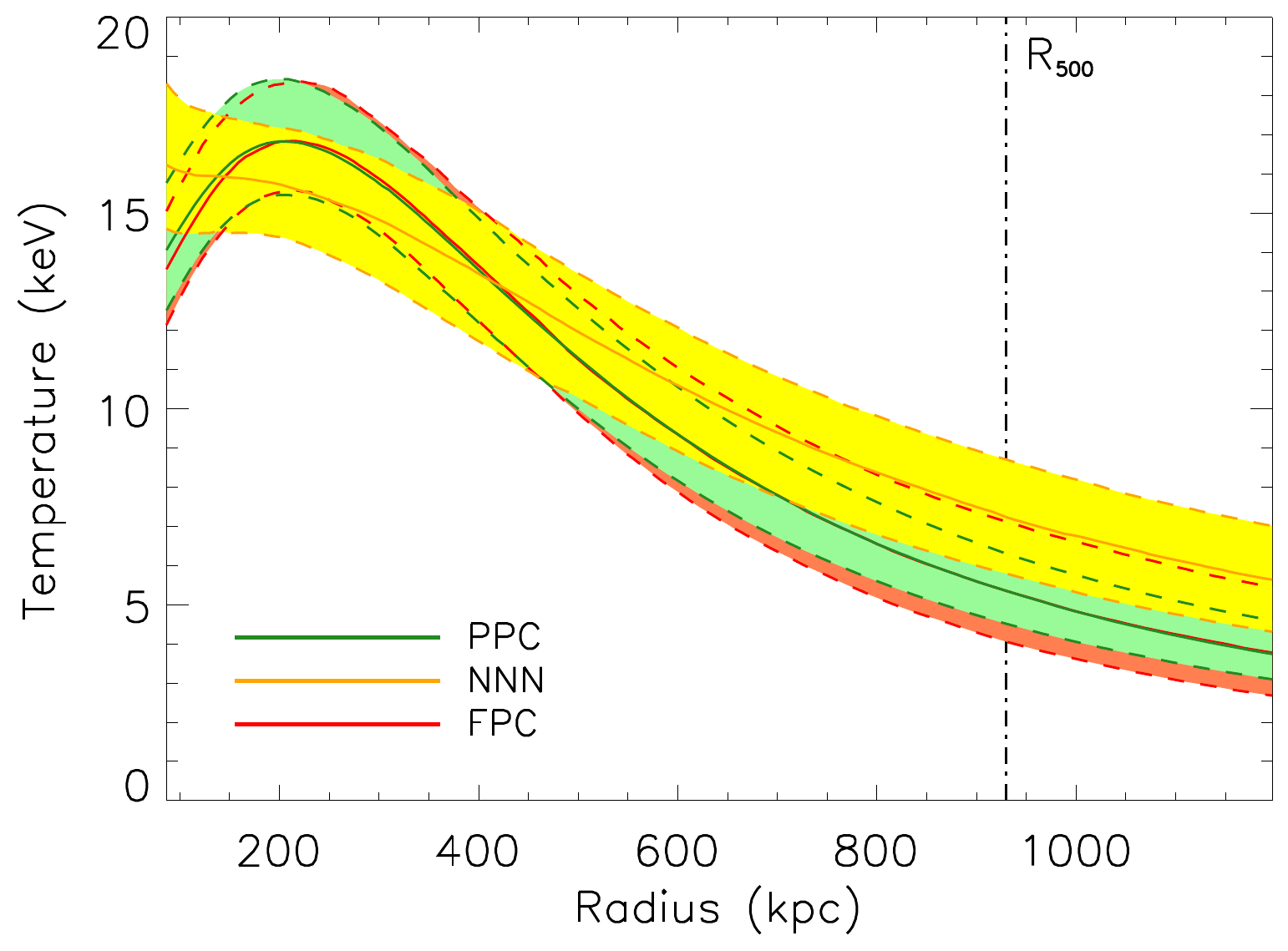}
	\includegraphics[height=4.3cm]{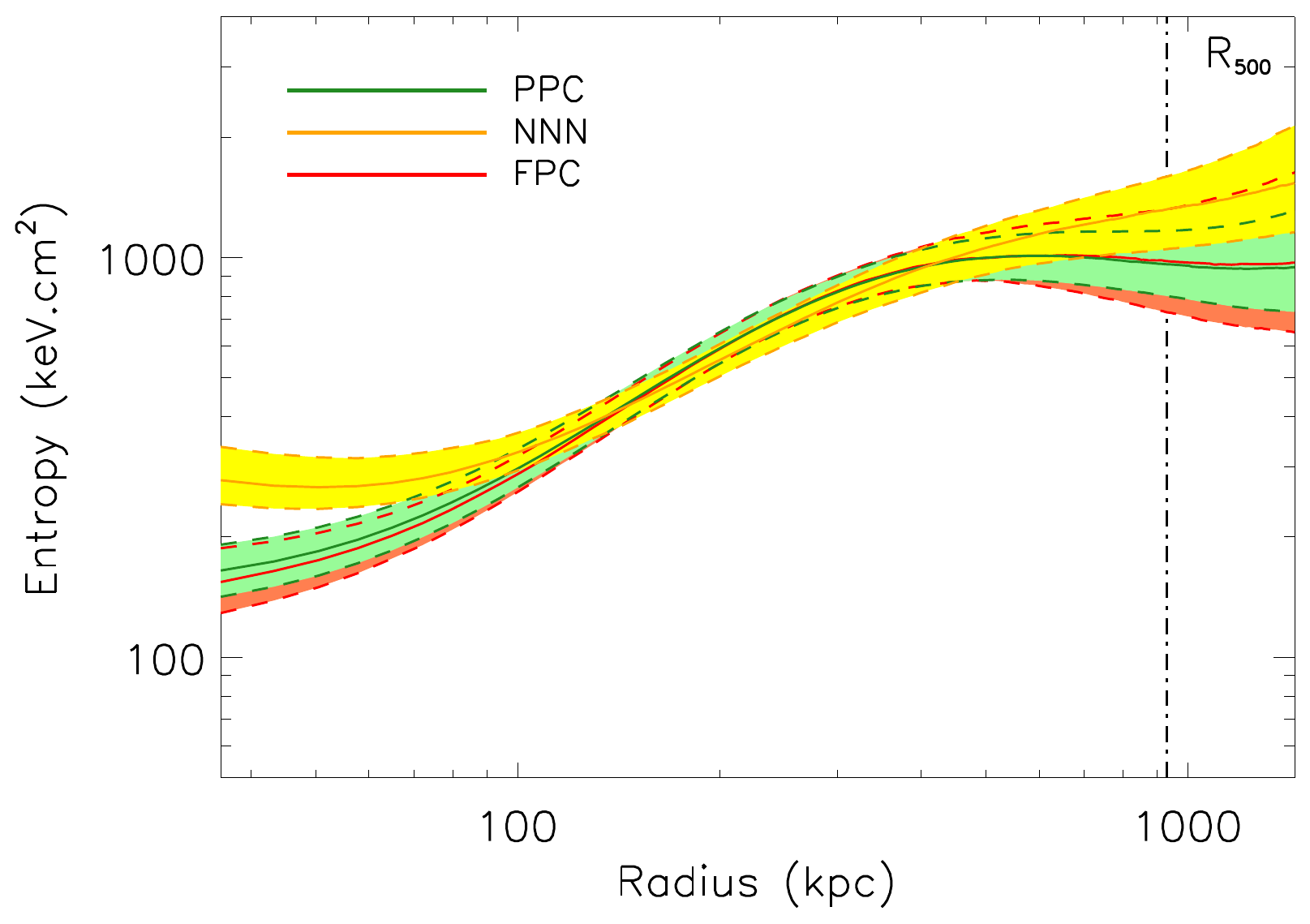}
	\caption{Pressure (left), temperature (middle) and entropy (right) profiles as a function of physical distance from the cluster center. The green, yellow, and red swaths provide the 68\% confidence limit, accounting for both calibration and statistical uncertainties. They correspond to modeling of the pressure profile with different choices for slope parameters $\left(a,b,c\right)$ as written in the legend. Once projected, a distance of 500~kpc corresponds to about 1 arcmin at the cluster redshift.}
        \label{fig:thermo_profile}
	\end{figure*}

%---------- Density
The best-fit density profile is represented in Fig.~\ref{fig:density_profile} with the data points of the ACCEPT catalog used to fit it. As shown in the bottom residual profile ($\chi$, the difference between data points and the best-fit model normalized by the data point errors), the model fits the data over the full radial range for the purpose of this paper. Since the best-fit density profile depends on the choice of the pressure profile slope only through the relativistic correction (see Sec.~\ref{sec:param_estim}), we show the result only for the first case, our baseline PPC pressure profile, and the differences between models are insignificant.

%---------- NIKA derived results
In Fig.~\ref{fig:thermo_profile} we present the radial distributions of the pressure, temperature, and entropy of the ICM of \mbox{CL~J1226.9+3332} derived using NIKA data. Uncertainties are given at 68\% confidence level and account for both statistical and NIKA overall calibration errors. The profiles corresponding to the different pressure profile models, PPC, NNN, and FPC (see Sect.~\ref{sec:icm_param_p}) are given.

%----------Pressure
The pressure (left) is characterized well by NIKA, with less than 10\% uncertainty below 500~kpc and up to 25\% at 1500~kpc for PPC. The profile is best constrained around 250~kpc, corresponding to $\sim$30 arcsec when projected onto the sky, where NIKA is most sensitive. The PPC profile is in qualitative agreement with the one obtained by \cite{mroczkowski2009}, despite a different choice of slope parameter. When using the same modeling, NNN, we find a good agreement, particularly on small scales. As expected, when loosening the constraints on parameter $a$, as in the case of the FPC profile, uncertainties increase by a factor of about 1.5. Thanks to the degeneracy between the pressure profile parameters, the different models agree well in the region where NIKA is sensitive. At both larger and smaller scales, the models tend to deviate from one another up to more than 1 $\sigma$, in particular NNN versus PPC and FPC, which directly propagates onto the other derived profiles, as discussed below.

%---------- Temperature
The temperature profile, derived from the pressure and the density, presents a core value of about 15~keV and decreases toward the outskirts of the cluster, reaching about 5~keV around 1500~kpc. For the PPC model we find uncertainties of about 10--15\%. The profile is slightly higher, but compatible within errors to those measured by \cite{mroczkowski2009} with SZA and \cite{maughan2007} with a detailed \mbox{X-ray} (Chandra + XMM) analysis. All three tested pressure profile models give compatible temperature results. Nevertheless, we notice that the core slope, $c$, obtained by \cite{comis2011} tends to indicate a cooler core below 200~kpc, while it is not the case for NNN. This is because the core temperature, which is the ratio between the pressure and the density, is directly related to the core pressure slope $c$. The flatter core of PPC and FPC, with respect to NNN, leads to a cooler core for the same density profile. Higher angular resolution tSZ observations would be necessary to provide constraints on the core slope pressure parameter $c$.

%---------- Entropy
The entropy profile is generally described well by $K(r)~=~K_0~+~K_{100}~\left(\frac{r}{100\ {\rm kpc}}\right)^{\alpha_K}$ \citep[e.g.,][]{pratt2010,cavagnolo2009}, where $K_0$ is called the core entropy, $K_{100}$ is a normalization, and $\alpha_K$ provides the slope of the profile. Large core entropies are expected for clusters with disturbed core such as \mbox{CL~J1226.9+3332}. The obtained entropy profile (right), shown on a logarithmic scale for both axes, is described well by a simple power law in the range directly probed by NIKA ($\gtrsim$~100~kpc). As we fixed the pressure profile parameter $c~=~0.014$ \citep{comis2011} in our baseline pressure model, which is expected to truly extrapolate the pressure parametrization on small scales, we expect the entropy profile to be accurate below100~kpc but limit ourselves to a qualitative discussion. The entropy profile shows signs of flattening below this scale with a core entropy above 100~keV cm$^2$, which would indicate that \mbox{CL~J1226.9+3332} is disturbed on small scales. At large radii, the slope seems to change but the error bars are too large for this effect to be measured. This discussion is also valid in the case of the NNN model, even if the core entropy tends to deviate by more than 1$\sigma$ between the two. Our baseline model, PPC, is fully compatible with the one obtained by \cite{maughan2007}. At large radii, however, the NNN profile is in better agreement with this \mbox{X-ray} analysis, with the main differences coming from the choice of the slope parameters at the core and the outer radii of the pressure profile.

%----------Mass distibution
	\begin{figure*}[h]
	\centering
	\includegraphics[height=6.3cm]{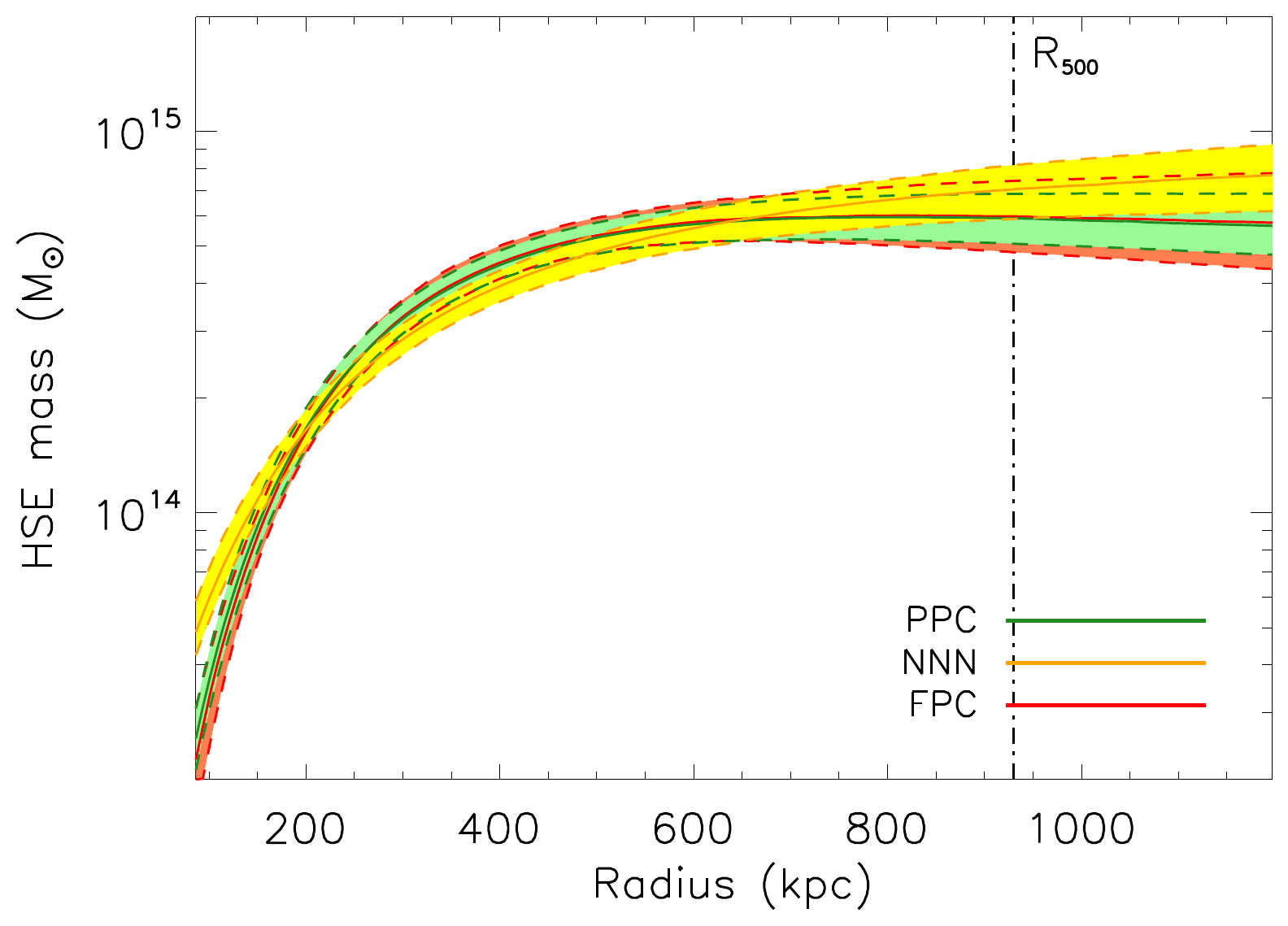}
	\includegraphics[height=6.3cm]{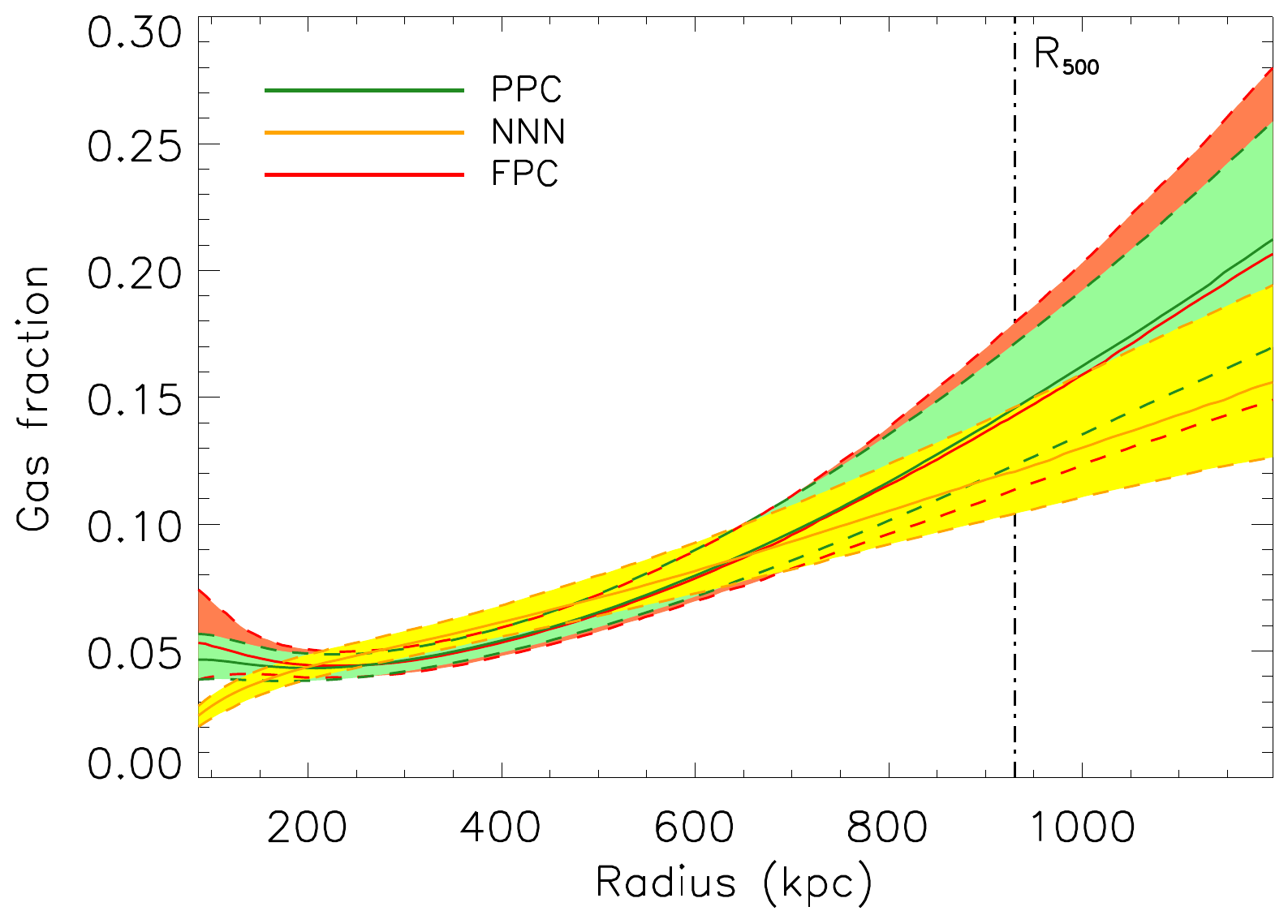}
	\caption{NIKA best-fit derived radial profiles for hydrostatic equilibrium (HSE) total mass (left) and the gas fraction (right). The color code is the same as in Fig.~\ref{fig:thermo_profile}.}
        \label{fig:mass_profile}
	\end{figure*}
The total mass and the gas fraction profiles are presented in Fig.~\ref{fig:mass_profile}. From the total mass profile we extract $R_{500}~=~930^{+50}_{-43}$ kpc, which in turn gives $M_{500}~=~5.96^{+1.02}_{-0.79}~\times~10^{14}$~M$_{\sun}$, compatible with previous measurements \citep[e.g.,][]{mroczkowski2009, maughan2007}. We obtain a gas fraction within $R_{500}$ of $f_{gas}(R_{500})~=~0.146^{+0.041}_{-0.030}$. The total mass PPC and NNN profiles give compatible results over the full radial range. Small differences between the two are most noticeable in the range where NIKA is not very sensitive, {\it i.e.}, on scales below 100~kpc where NNN is higher than PPC, mainly due to the difference in the pressure profile. The FPC model is fully compatible with the two other ones, and it presents larger error contours. The results are similar for the gas fraction, for which PPC presents a flattening below 200~kpc, while NNN keeps decreasing. Assuming the gas fraction of \mbox{CL~J1226.9+3332} within $R_{500}$ to be a good representation of the matter content in the Universe, we compare it to its expected gas fraction using \cite{planck2013param} cold dark matter, $\Omega_c$, and baryon density, $\Omega_b$, as $f_{gas} = \frac{\Omega_b}{\Omega_c + \Omega_b} = 0.156$. We find that it is compatible with our result within error bars.

%---------- Planck constraint compatibility and summary
The posterior ICM distribution is compatible in all cases with the integrated tSZ flux measured by Planck. The main outcomes of our analysis are summarized in Table~\ref{tab:results}. 
\begin{table}
\caption{Main results of the MCMC analysis. The quoted errors are given at 68\% confidence level.}
\begin{center}
\begin{tabular}{ccc}
\hline
\hline
PPC pressure profile \\
\hline
$M_{500}$ & $5.96^{+1.02}_{-0.79} \times 10^{14} \ M_{\sun}$ \\
$R_{500}$ & $930^{+50}_{-43}$ kpc\\
$\theta_{500}$ & $1.93^{+0.10}_{-0.09}$ arcmin\\
$f_{gas}(R_{500})$ & $0.146^{+0.041}_{-0.030}$\\
$Y_{500}$ & $0.598^{+0.063}_{-0.060} \times 10^{-3}$ arcmin$^2$\\
\hline
FPC pressure profile \\
\hline
$M_{500}$ & $6.10^{+1.52}_{-1.06} \times 10^{14} \ M_{\sun}$ \\
$R_{500}$ & $937^{+72}_{-58}$ kpc\\
$\theta_{500}$ & $1.95^{+0.15}_{-0.12}$ arcmin\\
$f_{gas}(R_{500})$ & $0.144^{+0.062}_{-0.038}$\\
$Y_{500}$ & $0.603^{+0.098}_{-0.070} \times 10^{-3}$ arcmin$^2$\\
\hline
NNN pressure profile \\
\hline
$M_{500}$ & $7.30^{+1.52}_{-1.34} \times 10^{14} \ M_{\sun}$ \\
$R_{500}$ & $995^{+65}_{-65}$ kpc\\
$\theta_{500}$ & $2.07^{+0.13}_{-0.13}$ arcmin\\
$f_{gas}(R_{500})$ & $0.129^{+0.041}_{-0.025}$\\
$Y_{500}$ & $0.717^{+0.117}_{-0.095} \times 10^{-3}$ arcmin$^2$\\
\hline
Point source PS260 &\\
\hline
150~GHz flux &  1.9 $\pm 0.2$ (stat.) $\pm 0.1$ (cal.) mJy\\
260~GHz flux & 6.8 $\pm 0.7$ (stat.) $\pm 1.0$ (cal.) mJy\\
260~GHz best-fit position R.A. & 12h~27m~0.01s\\
260~GHz best-fit position Dec. & 33$^o$~32'~42.0"\\
\hline
\end{tabular}
\end{center}
\label{tab:results}
\end{table}

%========== Scaling relations
\subsection{\mbox{CL~J1226.9+3332} and the tSZ--Mass scaling relations}\label{sec:scaling_relation}
	\begin{figure*}[h]
	\centering
	\includegraphics[height=6.3cm]{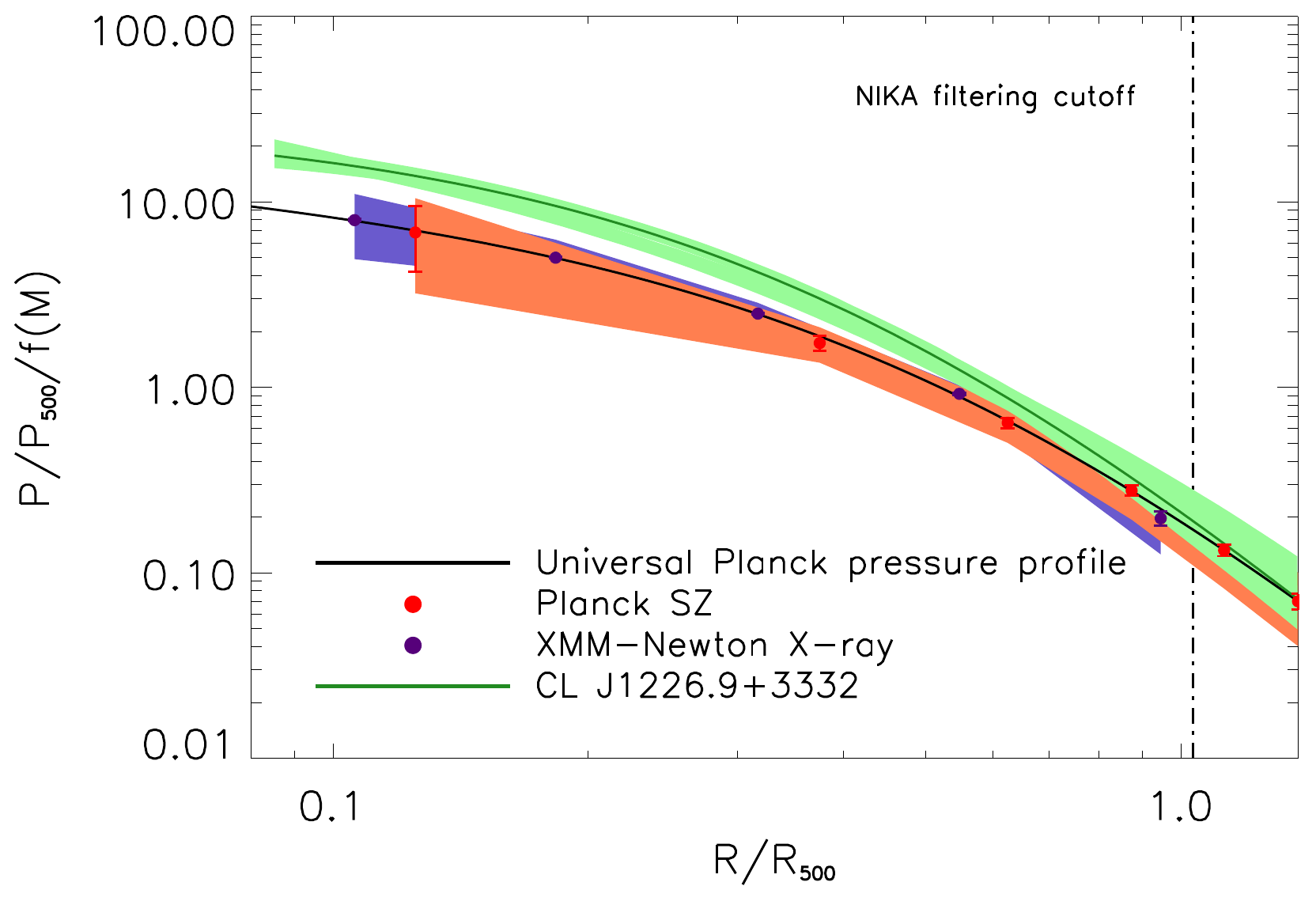}
	\includegraphics[height=6.3cm]{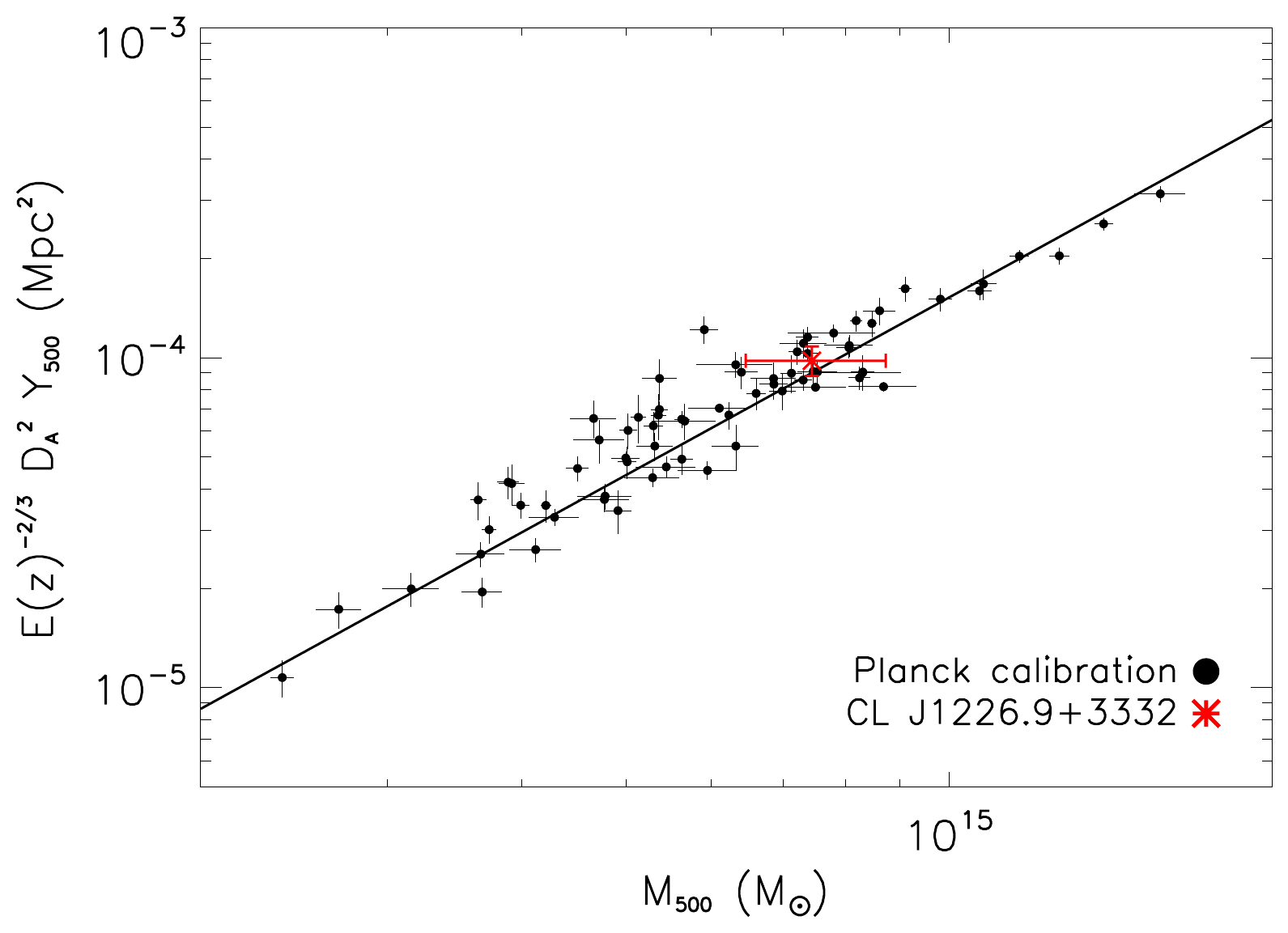}
	\caption{Left: Planck Universal pressure profile (black line), together with the best-fit profile obtained for \mbox{CL~J1226.9+3332} in green with the 1$\sigma$ error as a light green shadow accounting for all PPC, FPC, and NNN profile's. The Planck average of the individual pressure profiles across the 62 nearby cluster sample is given as red data points and the stacked pressure profile derived from the XMM data for the same sample is also given as purple dots \citep{planck2013pressure_profile}. The dispersion about the respective tSZ and \mbox{X-ray} profiles are shown by shaded area with similar (lighter) colors. The scale corresponding to the one at which the NIKA data start to be affected by filtering is also given as a vertical dashed line.
	Right: Planck $Y_{500}$--$M_{500}$ calibration \citep{planck2013cluster_count}, together with the NIKA value obtained for \mbox{CL~J1226.9+3332} in the case of the PPC pressure profile parameterization. The Planck scaling law is represented as a black line, and the data points of the clusters used for its calibration are given as black dots. The NIKA data point is given by the red star.}
        \label{fig:scaling_law}
	\end{figure*}
Clusters of galaxies are usually used for cosmological studies assuming a self-similar scenario because they are expected to be a scaled version of one another. In practice, non-gravitational processes can induce dispersion in the general trend and biases. The Planck satellite has recently released the largest tSZ selected cluster sample \citep[1227 objects,][]{planck2013catalogue}. To use this catalog for cosmology, \cite{planck2013cluster_count} have calibrated the relation between $Y_{\theta_{500}} \equiv Y_{500}$ and $M_{500}$ 
\begin{equation}
	E(z)^{-2/3} \left[\frac{D_A^2 Y_{500}}{10^{-4} {\rm Mpc}^2}\right] = 10^{-0.19 \pm 0.02} \left[\frac{(1-b) M_{500}}{6 \times 10^{14} M_{\sun}}\right]^{1.79 \pm 0.08},
\label{eq:Y500M500}
\end{equation}
where $E(z) = \sqrt{(1+z)^3 \Omega_M + \Omega_{\Lambda}}$. The extra bias term, $(1-b)$, corresponds to the expectation that the hydrostatic equilibrium (HSE) mass underestimates the true mass due to non-thermal pressure, so that $M^{HSE}_{500} = (1-b) M_{500}$. In this paper we set $b=0.2$, because it was the baseline for \cite{planck2013cluster_count}.

%---------- CL J1226 in this scaling law
As a demonstration of the potential of future NIKA2 tSZ dedicated large programs, we present a comparison of the recovered characteristics for \mbox{CL~J1226.9+3332} in terms of pressure profile and tSZ--mass scaling relation, to the Planck 2013 results \citep{planck2013pressure_profile,planck2013cluster_count}. The left panel of Fig.~\ref{fig:scaling_law} provides a comparison between the pressure profile of \mbox{CL~J1226.9+3332}, at high-redshift, and the average profile over 62 nearby clusters obtained by \cite{planck2013pressure_profile}. Both have been normalized to account for the mass and redshift dependance by $f(M) = \left(\frac{M_{500}}{3 \times 10^{14} \ M_{\sun}} \frac{H_0}{70 \ {\rm km s}^{-1} {\rm Mpc}^{-1}}\right)^{0.12}$ as detailed in~\cite{planck2013pressure_profile}. The NIKA data show that the normalized pressure profile of \mbox{CL~J1226.9+3332} is among the highest ones, but does not show any significant evidence of non-standard redshift evolution, within error bars. In addition we notice that our error bars are model dependent and do not reflect the full uncertainty of the data. The evolution of the pressure profile with redshift has been statistically tested recently using a Chandra \mbox{X-ray} analysis of 80 SPT clusters \citep{macdonald2014} with a highest bin at a mean redshift $z = 0.82$. They find no significant evolution, apart from the cluster's cores, and agree with a standard redshift evolution of the pressure distribution among clusters. In the right hand panel of Fig.~\ref{fig:scaling_law}, we present $Y_{500}$ as a function of $M_{500}$ for  \mbox{CL~J1226.9+3332}. For comparison we also show the best-fit \citet{planck2013cluster_count} scaling law and the data corresponding to the 71 clusters used for its calibration. The mean redshift of this cluster sample is 0.195 with a maximum redshift of 0.447. The cluster \mbox{CL~J1226.9+3332}, at $z=0.89$, is consistent with the \citet{planck2013cluster_count} scaling relation. This single data point does not allow us to draw any conclusion on the evolution with redshift. However, our results illustrate the strength of such measurements based on a sample of a few tens of clusters with future NIKA2 observations. This will indeed allow us to precisely constrain the redshift evolution of scaling relations, on the basis of individual measurements .

%###############################################################################################
%##########################                               CONCLUSION                                ##########################%###############################################################################################
\section{Summary and conclusions}\label{sec:conclusion}
%---------- Imaging of the cluster
The NIKA camera at the IRAM 30-meter telescope was used to image the cluster of galaxies \mbox{CL~J1226.9+3332} via the tSZ effect at 150 and 260~GHz with 18.2 and 12.0 arcsec angular resolution, respectively. It provides the first resolved observation of this cluster at these frequencies. The cluster signal is detected in the two bands, but our tSZ analysis focuses on the 150~GHz map since the signal-to-noise is higher at this frequency. A submillimeter point source is detected at 260~GHz, 30 arcsec away from the cluster center, showing the interest of the dual-band capabilities of NIKA to account for such contaminant. These observations, on scales $\sim$ 20 -- 200 arcsec, complement previous single-dish 90~GHz MUSTANG observations on scales in the range $\sim$ 10 -- 50 arcsec and interferometric SZA data at 30 and 90~GHz, which are the most sensitive at arcmin scales. The ICM morphology of the cluster agrees with these previous measurements. \mbox{CL~J1226.9+3332} appears relaxed on large scales and shows evidence of a disturbed core, most likely due to the merger of a subcluster to the southwest. It is also consistent with \mbox{X-ray} and lensing observations.

%---------- MCMC analysis
We used maximum likelihood analysis to constrain the pressure profile of the cluster via Markov Chain Monte Carlo sampling. The NIKA maps were combined with Chandra \mbox{X-ray} data using the ACCEPT data, to jointly derive ICM thermodynamic quantities (pressure, density, temperature, and entropy profiles). Planck tSZ data were also used to cross-check the overall flux of \mbox{CL~J1226.9+3332}. The inferred temperature profile of the cluster exhibits a core value of about 15~keV and decreases toward the outskirts, reaching about 5~keV around 1~Mpc. The entropy profile is described well by a simple power law in the range probed by NIKA but shows sign of flattening below 100~kpc with a core entropy above 100~keV cm$^2$, agreeing with \mbox{CL~J1226.9+3332} being disturbed on small scales. Assuming that the hydrostatic equilibrium accurately applies to this cluster, we extracted the total mass and gas mass profile and derived the gas fraction profile. We measured $R_{500}~=~930^{+50}_{-43}$ kpc and $M_{500}~=~5.96^{+1.02}_{-0.79}~\times~10^{14}$ M$_{\sun}$ at a 68\% confidence level. We compared these results when assuming Planck tSZ-based pressure-profile slope parameters or \mbox{X-ray}/numerical simulation based ones and find that both choices give consistent results in general. These results are compatible within the uncertainties with previous tSZ and \mbox{X-ray} measurements. 

%---------- NIKA2 perspectives
NIKA is the prototype of NIKA2, which will be permanently installed at the IRAM 30-meter telescope at the end of 2015. NIKA2 will contain about 5000 detectors, {\it i.e.}, 15 times more than NIKA, within the same frequency bands and similar angular resolution. Its instantaneous field of view will accordingly increase from 1.8 to 6.5 arcmin. With such characteristics, NIKA2 will be well adapted to mapping the tSZ signal in intermediate and distant clusters of galaxies. The observation of \mbox{CL~J1226.9+3332} is part of a pilot study that aims at characterizing the possible scientific outcomes of large observing campaigns with NIKA2. Future NIKA2 dedicated tSZ observations of a few tens of clusters would allow study of the evolution of scaling and structural properties of clusters of galaxies out to $z \sim 1$. Here, by comparing our results to the expected tSZ--mass Planck scaling relations for a single cluster, we have shown that with more objects, NIKA2 will be able to calibrate the tSZ-mass scaling relation and its eventual redshift dependence. 

%###############################################################################################
%##########################                       ACKNOWLEDGEMENTS                        ##########################%###############################################################################################
\begin{acknowledgements}
We are thankful to the anonymous referee for useful comments that helped improve the quality of the paper.
We gratefully thank Marian Douspis and the Planck collaboration for providing the Planck data points shown in the right hand panel of Fig.~\ref{fig:scaling_law}.
We thank Marco De Petris for useful comments.
We would like to thank the IRAM staff for their support during the campaign.
This work has been partially funded by the Foundation Nanoscience Grenoble, the ANR under the contracts "MKIDS" and "NIKA". 
This work has been partially supported by the LabEx FOCUS ANR-11-LABX-0013. 
This work has benefited from the support of the European Research Council Advanced Grant ORISTARS under the European Union's Seventh Framework Program (Grant Agreement no. 291294).
The NIKA dilution cryostat was designed and built at the Institut N\'eel. In particular, we acknowledge the crucial contribution of the Cryogenics Group and, in particular Gregory Garde, Henri Rodenas, Jean Paul Leggeri, and Philippe Camus. 
R. A. would like to thank the ENIGMASS French LabEx for funding this work. 
B. C. acknowledges support from the CNES post-doctoral fellowship program. 
E. P. acknowledges support from grant ANR-11-BS56-015. 
A. R. would like to thank the FOCUS French LabEx for funding this work.
A. R. acknowledges support from the CNES doctoral fellowship program.
\end{acknowledgements}

\bibliography{biblio}
\end{document}

%% file: listeauthors.tex
\author{R.~Adam\inst{\ref{inst1}}
\and B.~Comis\inst{\ref{inst1}}
\and J.-F.~Mac\'ias-P\'erez\inst{\ref{inst1}}
\and A.~Adane\inst{\ref{inst2}}
\and P.~Ade\inst{\ref{inst3}}
\and P.~Andr\'e\inst{\ref{inst4}}
\and A.~Beelen\inst{\ref{inst5}}
\and B.~Belier\inst{\ref{inst6}}
\and A.~Beno\^it\inst{\ref{inst7}}
\and A.~Bideaud\inst{\ref{inst3}}
\and N.~Billot\inst{\ref{inst8}}
\and G.~Blanquer\inst{\ref{inst1}}
\and O.~Bourrion\inst{\ref{inst1}}
\and M.~Calvo\inst{\ref{inst7}}
\and A.~Catalano\inst{\ref{inst1}}
\and G.~Coiffard\inst{\ref{inst2}}
\and A.~Cruciani\inst{\ref{inst14}}
\and A.~D'Addabbo\inst{\ref{inst7}, \ref{inst14}}
\and F.-X.~D\'esert\inst{\ref{inst9}}
\and S.~Doyle\inst{\ref{inst3}}
\and J.~Goupy\inst{\ref{inst7}}
\and C.~Kramer\inst{\ref{inst8}}
\and S.~Leclercq\inst{\ref{inst2}}
\and J.~Martino\inst{\ref{inst5}}
\and P.~Mauskopf\inst{\ref{inst3}, \ref{inst13}}
\and F.~Mayet\inst{\ref{inst1}}
\and A.~Monfardini\inst{\ref{inst7}}
\and F.~Pajot\inst{\ref{inst5}}
\and E.~Pascale\inst{\ref{inst3}}
\and L.~Perotto\inst{\ref{inst1}}
\and E.~Pointecouteau\inst{\ref{inst10}, \ref{inst11}}
\and N.~Ponthieu\inst{\ref{inst9}}
\and V.~Rev\'eret\inst{\ref{inst4}}
\and A.~Ritacco\inst{\ref{inst1}}
\and L.~Rodriguez\inst{\ref{inst4}}
\and G.~Savini\inst{\ref{inst12}}
\and K.~Schuster\inst{\ref{inst2}}
\and A.~Sievers\inst{\ref{inst8}}
\and C.~Tucker\inst{\ref{inst3}}
\and R.~Zylka\inst{\ref{inst2}}}

\offprints{R. Adam - adam@lpsc.in2p3.fr}

\institute{
Laboratoire de Physique Subatomique et de Cosmologie (LPSC), Universit\'e Grenoble-Alpes, CNRS/IN2P3, 53 avenue des Martyrs, 38026 Grenoble, France
  \label{inst1}
\and
Institut de RadioAstronomie Millim\'etrique (IRAM), Grenoble, France
  \label{inst2}
\and
Astronomy Instrumentation Group, University of Cardiff, UK
  \label{inst3}
\and
Laboratoire AIM, CEA/IRFU, CNRS/INSU, Universit\'e Paris Diderot, CEA-Saclay, 91191 Gif-Sur-Yvette, France 
  \label{inst4}
\and
Institut d'Astrophysique Spatiale (IAS), CNRS and Universit\'e Paris Sud, Orsay, France
  \label{inst5}
\and
Institut d'Electronique Fondamentale (IEF), Universit\'e Paris Sud, Orsay, France
  \label{inst6}
\and
Institut N\'eel, CNRS and Universit\'e de Grenoble, France
  \label{inst7}
\and
Institut de RadioAstronomie Millim\'etrique (IRAM), Granada, Spain
  \label{inst8}
\and
Univ. Grenoble Alpes, IPAG, F-38000 Grenoble, France\\
CNRS, IPAG, F-38000 Grenoble, France 
  \label{inst9}
\and
Universit\'e de Toulouse, UPS-OMP, Institut de Recherche en Astrophysique et Plan\'etologie (IRAP), Toulouse, France
  \label{inst10}
\and
CNRS, IRAP, 9 Av. colonel Roche, BP 44346, F-31028 Toulouse cedex 4, France 
  \label{inst11}
\and
University College London, Department 
of Physics and Astronomy, Gower Street, London WC1E 6BT, UK
  \label{inst12}
  \and
  School of Earth and Space Exploration and Department of Physics, 
	Arizona State University, Tempe, AZ 85287
  \label{inst13}
    \and
  Dipartimento di Fisica, Sapienza Universit\`a di Roma, Piazzale Aldo Moro 5, I-00185 Roma, Italy
  \label{inst14}
}

%% file: NIKA_CL1227.bbl
\begin{thebibliography}{54}
\expandafter\ifx\csname natexlab\endcsname\relax\def\natexlab#1{#1}\fi

\bibitem[{{Adam} {et~al.}(2014){Adam}, {Comis}, {Mac{\'{\i}}as-P{\'e}rez},
  {Adane}, {Ade}, {Andr{\'e}}, {Beelen}, {Belier}, {Beno{\^i}t}, {Bideaud},
  {Billot}, {Boudou}, {Bourrion}, {Calvo}, {Catalano}, {Coiffard}, {D'Addabbo},
  {D{\'e}sert}, {Doyle}, {Goupy}, {Kramer}, {Leclercq}, {Martino}, {Mauskopf},
  {Mayet}, {Monfardini}, {Pajot}, {Pascale}, {Perotto}, {Pointecouteau},
  {Ponthieu}, {Rev{\'e}ret}, {Rodriguez}, {Savini}, {Schuster}, {Sievers},
  {Tucker}, \& {Zylka}}]{adam2013}
{Adam}, R. {et~al.} 2014, \aap, 569, A66, 1310.6237

\bibitem[{{Allen} {et~al.}(2011){Allen}, {Evrard}, \& {Mantz}}]{allen2011}
{Allen}, S.~W., {Evrard}, A.~E., \& {Mantz}, A.~B. 2011, \araa, 49, 409,
  1103.4829

\bibitem[{{Arnaud} {et~al.}(2010){Arnaud}, {Pratt}, {Piffaretti},
  {B{\"o}hringer}, {Croston}, \& {Pointecouteau}}]{arnaud2010}
{Arnaud}, M., {Pratt}, G.~W., {Piffaretti}, R., {B{\"o}hringer}, H., {Croston},
  J.~H., \& {Pointecouteau}, E. 2010, \aap, 517, A92, 0910.1234

\bibitem[{{Bartelmann}(2010)}]{bartelmann2010}
{Bartelmann}, M. 2010, Classical and Quantum Gravity, 27, 233001, 1010.3829

\bibitem[{{Birkinshaw}(1999)}]{birkinshaw1999}
{Birkinshaw}, M. 1999, \physrep, 310, 97, arXiv:astro-ph/9808050

\bibitem[{{Bleem} {et~al.}(2014){Bleem}, {Stalder}, {de Haan}, {Aird}, {Allen},
  {Applegate}, {Ashby}, {Bautz}, {Bayliss}, {Benson}, {Bocquet}, {Brodwin},
  {Carlstrom}, {Chang}, {Chiu}, {Cho}, {Clocchiatti}, {Crawford}, {Crites},
  {Desai}, {Dietrich}, {Dobbs}, {Foley}, {Forman}, {George}, {Gladders},
  {Gonzalez}, {Halverson}, {Hennig}, {Hoekstra}, {Holder}, {Holzapfel},
  {Hrubes}, {Jones}, {Keisler}, {Knox}, {Lee}, {Leitch}, {Liu}, {Lueker},
  {Luong-Van}, {Mantz}, {Marrone}, {McDonald}, {McMahon}, {Meyer}, {Mocanu},
  {Mohr}, {Murray}, {Padin}, {Pryke}, {Reichardt}, {Rest}, {Ruel}, {Ruhl},
  {Saliwanchik}, {Saro}, {Sayre}, {Schaffer}, {Schrabback}, {Shirokoff},
  {Song}, {Spieler}, {Stanford}, {Staniszewski}, {Stark}, {Story}, {Stubbs},
  {Vanderlinde}, {Vieira}, {Vikhlinin}, {Williamson}, {Zahn}, \&
  {Zenteno}}]{bleem2014}
{Bleem}, L.~E. {et~al.} 2014, ArXiv e-prints, 1409.0850

\bibitem[{{B{\"o}hringer} \& {Werner}(2010)}]{bohringer2010}
{B{\"o}hringer}, H., \& {Werner}, N. 2010, \aapr, 18, 127

\bibitem[{{Bonamente} {et~al.}(2006){Bonamente}, {Joy}, {LaRoque}, {Carlstrom},
  {Reese}, \& {Dawson}}]{bonamente2006}
{Bonamente}, M., {Joy}, M.~K., {LaRoque}, S.~J., {Carlstrom}, J.~E., {Reese},
  E.~D., \& {Dawson}, K.~S. 2006, \apj, 647, 25, astro-ph/0512349

\bibitem[{{Bourrion} {et~al.}(2011){Bourrion}, {Bideaud}, {Benoit}, {Cruciani},
  {Macias-Perez}, {Monfardini}, {Roesch}, {Swenson}, \&
  {Vescovi}}]{bourion2011}
{Bourrion}, O. {et~al.} 2011, JINST, 6, P06012, 1102.1314

\bibitem[{{Calvo} {et~al.}(2012){Calvo}, {Roesch}, {D{\'e}sert}, {Benoit},
  {Monfardini}, {Ade}, {Boudou}, {Bourrion}, {Camus}, {Cruciani}, {Doyle},
  {Hoffmann}, {Leclercq}, {Macias-Perez}, {Mauskopf}, {Ponthieu}, {Schuster},
  {Tucker}, \& {Vescovi}}]{calvo2012}
{Calvo}, M. {et~al.} 2012, \aap

\bibitem[{{Carlstrom} {et~al.}(2002){Carlstrom}, {Holder}, \&
  {Reese}}]{carlstrom2002}
{Carlstrom}, J.~E., {Holder}, G.~P., \& {Reese}, E.~D. 2002, \araa, 40, 643,
  arXiv:astro-ph/0208192

\bibitem[{{Catalano} {et~al.}(2014){Catalano}, {Calvo}, {Ponthieu}, {Adam},
  {Adane}, {Ade}, {Andre}, {Beelen}, {Belier}, {Benoit}, {Bideaud}, {Billot},
  {Boudou}, {Bourrion}, {Coiffard}, {Comis}, {D'Addabbo}, {Desert}, {Doyle},
  {Goupy}, {Kramer}, {Leclercq}, {Macias-Perez}, {Martino}, {Mauskopf},
  {Mayet}, {Monfardini}, {Pajot}, {Pascale}, {Perotto}, {Reveret}, {Rodriguez},
  {Savini}, {Schuster}, {Sievers}, {Tucker}, \& {Zylka}}]{catalano2014}
{Catalano}, A. {et~al.} 2014, ArXiv e-prints, 1402.0260

\bibitem[{{Cavagnolo} {et~al.}(2009){Cavagnolo}, {Donahue}, {Voit}, \&
  {Sun}}]{cavagnolo2009}
{Cavagnolo}, K.~W., {Donahue}, M., {Voit}, G.~M., \& {Sun}, M. 2009, \apjs,
  182, 12, 0902.1802

\bibitem[{{Cavaliere} \& {Fusco-Femiano}(1978)}]{cavaliere1978}
{Cavaliere}, A., \& {Fusco-Femiano}, R. 1978, \aap, 70, 677

\bibitem[{Chib \& Greenberg(1995)}]{chib1995}
Chib, S., \& Greenberg, E. 1995, The American Statistician, 49, 327

\bibitem[{{Comis} {et~al.}(2011){Comis}, {de Petris}, {Conte}, {Lamagna}, \&
  {de Gregori}}]{comis2011}
{Comis}, B., {de Petris}, M., {Conte}, A., {Lamagna}, L., \& {de Gregori}, S.
  2011, \mnras, 418, 1089, 1108.1029

\bibitem[{{Ebeling} {et~al.}(2001){Ebeling}, {Jones}, {Fairley}, {Perlman},
  {Scharf}, \& {Horner}}]{ebeling2001}
{Ebeling}, H., {Jones}, L.~R., {Fairley}, B.~W., {Perlman}, E., {Scharf}, C.,
  \& {Horner}, D. 2001, \apjl, 548, L23, astro-ph/0012175

\bibitem[{{Feretti} {et~al.}(2011){Feretti}, {Giovannini}, {Govoni}, \&
  {Murgia}}]{feretti2011}
{Feretti}, L., {Giovannini}, G., {Govoni}, F., \& {Murgia}, M. 2011, in IAU
  Symposium, Vol. 274, IAU Symposium, ed. A.~{Bonanno}, E.~{de Gouveia Dal
  Pino}, \& A.~G. {Kosovichev}, 340--347, 1101.1887

\bibitem[{Gelman \& Rubin(1992)}]{gelman1992}
Gelman, A., \& Rubin, D.~B. 1992, Statistical Science, 7, 457

\bibitem[{{Hasselfield} {et~al.}(2013){Hasselfield}, {Hilton}, {Marriage},
  {Addison}, {Barrientos}, {Battaglia}, {Battistelli}, {Bond}, {Crichton},
  {Das}, {Devlin}, {Dicker}, {Dunkley}, {D{\"u}nner}, {Fowler}, {Gralla},
  {Hajian}, {Halpern}, {Hincks}, {Hlozek}, {Hughes}, {Infante}, {Irwin},
  {Kosowsky}, {Marsden}, {Menanteau}, {Moodley}, {Niemack}, {Nolta}, {Page},
  {Partridge}, {Reese}, {Schmitt}, {Sehgal}, {Sherwin}, {Sievers}, {Sif{\'o}n},
  {Spergel}, {Staggs}, {Swetz}, {Switzer}, {Thornton}, {Trac}, \&
  {Wollack}}]{hasselfield2013}
{Hasselfield}, M. {et~al.} 2013, \jcap, 7, 8, 1301.0816

\bibitem[{{Holden} {et~al.}(2009){Holden}, {Franx}, {Illingworth}, {Postman},
  {van der Wel}, {Kelson}, {Blakeslee}, {Ford}, {Demarco}, \&
  {Mei}}]{holden2009}
{Holden}, B.~P. {et~al.} 2009, \apj, 693, 617, 0811.1986

\bibitem[{{Hurier} {et~al.}(2013){Hurier}, {Mac{\'{\i}}as-P{\'e}rez}, \&
  {Hildebrandt}}]{hurier2013}
{Hurier}, G., {Mac{\'{\i}}as-P{\'e}rez}, J.~F., \& {Hildebrandt}, S. 2013,
  \aap, 558, A118, 1007.1149

\bibitem[{{Itoh} {et~al.}(1998){Itoh}, {Kohyama}, \& {Nozawa}}]{itoh1998}
{Itoh}, N., {Kohyama}, Y., \& {Nozawa}, S. 1998, \apj, 502, 7,
  arXiv:astro-ph/9712289

\bibitem[{{Jee} \& {Tyson}(2009)}]{jee2009}
{Jee}, M.~J., \& {Tyson}, J.~A. 2009, \apj, 691, 1337, 0810.0709

\bibitem[{{Joy} {et~al.}(2001){Joy}, {LaRoque}, {Grego}, {Carlstrom}, {Dawson},
  {Ebeling}, {Holzapfel}, {Nagai}, \& {Reese}}]{joy2001}
{Joy}, M. {et~al.} 2001, \apjl, 551, L1, astro-ph/0012052

\bibitem[{{Kitayama}(2014)}]{kitayama2014}
{Kitayama}, T. 2014, ArXiv e-prints, 1404.0870

\bibitem[{{Korngut} {et~al.}(2011){Korngut}, {Dicker}, {Reese}, {Mason},
  {Devlin}, {Mroczkowski}, {Sarazin}, {Sun}, \& {Sievers}}]{korngut2011}
{Korngut}, P.~M. {et~al.} 2011, \apj, 734, 10, 1010.5494

\bibitem[{{Maughan} {et~al.}(2007){Maughan}, {Jones}, {Jones}, \& {Van
  Speybroeck}}]{maughan2007}
{Maughan}, B.~J., {Jones}, C., {Jones}, L.~R., \& {Van Speybroeck}, L. 2007,
  \apj, 659, 1125, astro-ph/0609690

\bibitem[{{Maughan} {et~al.}(2004){Maughan}, {Jones}, {Ebeling}, \&
  {Scharf}}]{maughan2004}
{Maughan}, B.~J., {Jones}, L.~R., {Ebeling}, H., \& {Scharf}, C. 2004, \mnras,
  351, 1193, astro-ph/0403521

\bibitem[{{McDonald} {et~al.}(2014){McDonald}, {Benson}, {Vikhlinin}, {Aird},
  {Allen}, {Bautz}, {Bayliss}, {Bleem}, {Bocquet}, {Brodwin}, {Carlstrom},
  {Chang}, {Cho}, {Clocchiatti}, {Crawford}, {Crites}, {de Haan}, {Dobbs},
  {Foley}, {Forman}, {George}, {Gladders}, {Gonzalez}, {Halverson},
  {Hlavacek-Larrondo}, {Holder}, {Holzapfel}, {Hrubes}, {Jones}, {Keisler},
  {Knox}, {Lee}, {Leitch}, {Liu}, {Lueker}, {Luong-Van}, {Mantz}, {Marrone},
  {McMahon}, {Meyer}, {Miller}, {Mocanu}, {Mohr}, {Murray}, {Padin}, {Pryke},
  {Reichardt}, {Rest}, {Ruhl}, {Saliwanchik}, {Saro}, {Sayre}, {Schaffer},
  {Shirokoff}, {Spieler}, {Stalder}, {Stanford}, {Staniszewski}, {Stark},
  {Story}, {Stubbs}, {Vanderlinde}, {Vieira}, {Williamson}, {Zahn}, \&
  {Zenteno}}]{macdonald2014}
{McDonald}, M. {et~al.} 2014, ArXiv e-prints, 1404.6250

\bibitem[{{Monfardini} {et~al.}(2011){Monfardini}, {Benoit}, {Bideaud},
  {Swenson}, {Cruciani}, {Camus}, {Hoffmann}, {D{\'e}sert}, {Doyle}, {Ade},
  {Mauskopf}, {Tucker}, {Roesch}, {Leclercq}, {Schuster}, {Endo}, {Baryshev},
  {Baselmans}, {Ferrari}, {Yates}, {Bourrion}, {Macias-Perez}, {Vescovi},
  {Calvo}, \& {Giordano}}]{monfardini2011}
{Monfardini}, A. {et~al.} 2011, \apjs, 194, 24, 1102.0870

\bibitem[{{Monfardini} {et~al.}(2010){Monfardini}, {Swenson}, {Bideaud},
  {D{\'e}sert}, {Yates}, {Benoit}, {Baryshev}, {Baselmans}, {Doyle}, {Klein},
  {Roesch}, {Tucker}, {Ade}, {Calvo}, {Camus}, {Giordano}, {Guesten},
  {Hoffmann}, {Leclercq}, {Mauskopf}, \& {Schuster}}]{monfardini2010}
------. 2010, \aap, 521, A29, 1004.2209

\bibitem[{{Moreno}(2010)}]{moreno2010}
{Moreno}, R. 2010, Neptune and Uranus planetary brightness temperature
  tabulation. Tech. rep., ESA Herschel Science Center, available from
  ftp://ftp.sciops.esa.int/pub/hsc-calibration/PlanetaryModels/ESA2

\bibitem[{{Mroczkowski}(2011)}]{mroczkowski2011}
{Mroczkowski}, T. 2011, \apjl, 728, L35, 1101.2176

\bibitem[{{Mroczkowski} {et~al.}(2009){Mroczkowski}, {Bonamente}, {Carlstrom},
  {Culverhouse}, {Greer}, {Hawkins}, {Hennessy}, {Joy}, {Lamb}, {Leitch},
  {Loh}, {Maughan}, {Marrone}, {Miller}, {Muchovej}, {Nagai}, {Pryke}, {Sharp},
  \& {Woody}}]{mroczkowski2009}
{Mroczkowski}, T. {et~al.} 2009, \apj, 694, 1034, 0809.5077

\bibitem[{{Muchovej} {et~al.}(2007){Muchovej}, {Mroczkowski}, {Carlstrom},
  {Cartwright}, {Greer}, {Hennessy}, {Loh}, {Pryke}, {Reddall}, {Runyan},
  {Sharp}, {Hawkins}, {Lamb}, {Woody}, {Joy}, {Leitch}, \&
  {Miller}}]{muchovej2007}
{Muchovej}, S. {et~al.} 2007, \apj, 663, 708, astro-ph/0610115

\bibitem[{{Nagai}(2006)}]{nagai2006}
{Nagai}, D. 2006, \apj, 650, 538, astro-ph/0512208

\bibitem[{{Nagai} {et~al.}(2007{\natexlab{a}}){Nagai}, {Kravtsov}, \&
  {Vikhlinin}}]{nagai2007b}
{Nagai}, D., {Kravtsov}, A.~V., \& {Vikhlinin}, A. 2007{\natexlab{a}}, \apj,
  668, 1, astro-ph/0703661

\bibitem[{{Nagai} {et~al.}(2007{\natexlab{b}}){Nagai}, {Vikhlinin}, \&
  {Kravtsov}}]{nagai2007}
{Nagai}, D., {Vikhlinin}, A., \& {Kravtsov}, A.~V. 2007{\natexlab{b}}, \apj,
  655, 98, arXiv:astro-ph/0609247

\bibitem[{{Planck Collaboration} {et~al.}(2013{\natexlab{a}}){Planck
  Collaboration}, {Ade}, {Aghanim}, {Alves}, {Armitage-Caplan}, {Arnaud},
  {Ashdown}, {Atrio-Barandela}, {Aumont}, {Aussel}, \&
  et~al.}]{planck2013mission}
{Planck Collaboration} {et~al.} 2013{\natexlab{a}}, ArXiv e-prints, 1303.5062

\bibitem[{{Planck Collaboration} {et~al.}(2013{\natexlab{b}}){Planck
  Collaboration}, {Ade}, {Aghanim}, {Armitage-Caplan}, {Arnaud}, {Ashdown},
  {Atrio-Barandela}, {Aumont}, {Aussel}, {Baccigalupi}, \&
  et~al.}]{planck2013catalogue}
------. 2013{\natexlab{b}}, ArXiv e-prints, 1303.5089

\bibitem[{{Planck Collaboration} {et~al.}(2013{\natexlab{c}}){Planck
  Collaboration}, {Ade}, {Aghanim}, {Armitage-Caplan}, {Arnaud}, {Ashdown},
  {Atrio-Barandela}, {Aumont}, {Baccigalupi}, {Banday}, \&
  et~al.}]{planck2013calib}
------. 2013{\natexlab{c}}, ArXiv e-prints, 1303.5069

\bibitem[{{Planck Collaboration} {et~al.}(2013{\natexlab{d}}){Planck
  Collaboration}, {Ade}, {Aghanim}, {Armitage-Caplan}, {Arnaud}, {Ashdown},
  {Atrio-Barandela}, {Aumont}, {Baccigalupi}, {Banday}, \&
  et~al.}]{planck2013param}
------. 2013{\natexlab{d}}, ArXiv e-prints, 1303.5076

\bibitem[{{Planck Collaboration} {et~al.}(2013{\natexlab{e}}){Planck
  Collaboration}, {Ade}, {Aghanim}, {Armitage-Caplan}, {Arnaud}, {Ashdown},
  {Atrio-Barandela}, {Aumont}, {Baccigalupi}, {Banday}, \&
  et~al.}]{planck2013cluster_count}
------. 2013{\natexlab{e}}, ArXiv e-prints, 1303.5080

\bibitem[{{Planck Collaboration} {et~al.}(2013{\natexlab{f}}){Planck
  Collaboration}, {Ade}, {Aghanim}, {Armitage-Caplan}, {Arnaud}, {Ashdown},
  {Atrio-Barandela}, {Aumont}, {Baccigalupi}, {Banday}, \&
  et~al.}]{planck2013ymap}
------. 2013{\natexlab{f}}, ArXiv e-prints, 1303.5081

\bibitem[{{Planck Collaboration} {et~al.}(2013{\natexlab{g}}){Planck
  Collaboration}, {Ade}, {Aghanim}, {Arnaud}, {Ashdown}, {Atrio-Barandela},
  {Aumont}, {Baccigalupi}, {Balbi}, {Banday}, \&
  et~al.}]{planck2013pressure_profile}
------. 2013{\natexlab{g}}, \aap, 550, A131, 1207.4061

\bibitem[{{Pratt} {et~al.}(2010){Pratt}, {Arnaud}, {Piffaretti},
  {B{\"o}hringer}, {Ponman}, {Croston}, {Voit}, {Borgani}, \&
  {Bower}}]{pratt2010}
{Pratt}, G.~W. {et~al.} 2010, \aap, 511, A85, 0909.3776

\bibitem[{{Reichardt} {et~al.}(2013){Reichardt}, {Stalder}, {Bleem}, {Montroy},
  {Aird}, {Andersson}, {Armstrong}, {Ashby}, {Bautz}, {Bayliss}, {Bazin},
  {Benson}, {Brodwin}, {Carlstrom}, {Chang}, {Cho}, {Clocchiatti}, {Crawford},
  {Crites}, {de Haan}, {Desai}, {Dobbs}, {Dudley}, {Foley}, {Forman}, {George},
  {Gladders}, {Gonzalez}, {Halverson}, {Harrington}, {High}, {Holder},
  {Holzapfel}, {Hoover}, {Hrubes}, {Jones}, {Joy}, {Keisler}, {Knox}, {Lee},
  {Leitch}, {Liu}, {Lueker}, {Luong-Van}, {Mantz}, {Marrone}, {McDonald},
  {McMahon}, {Mehl}, {Meyer}, {Mocanu}, {Mohr}, {Murray}, {Natoli}, {Padin},
  {Plagge}, {Pryke}, {Rest}, {Ruel}, {Ruhl}, {Saliwanchik}, {Saro}, {Sayre},
  {Schaffer}, {Shaw}, {Shirokoff}, {Song}, {Spieler}, {Staniszewski}, {Stark},
  {Story}, {Stubbs}, {{\v S}uhada}, {van Engelen}, {Vanderlinde}, {Vieira},
  {Vikhlinin}, {Williamson}, {Zahn}, \& {Zenteno}}]{reichardt2013}
{Reichardt}, C.~L. {et~al.} 2013, \apj, 763, 127, 1203.5775

\bibitem[{{Sarazin}(1988)}]{sarazin1988}
{Sarazin}, C.~L. 1988, \skytel, 76, 639

\bibitem[{{Sunyaev} \& {Zel'dovich}(1972)}]{sunyaev1972}
{Sunyaev}, R.~A., \& {Zel'dovich}, Y.~B. 1972, \apspr, 4, 173

\bibitem[{{Sunyaev} \& {Zel'dovich}(1980)}]{sunyaev1980}
------. 1980, \araa, 18, 537

\bibitem[{{Vikhlinin} {et~al.}(2006){Vikhlinin}, {Kravtsov}, {Forman}, {Jones},
  {Markevitch}, {Murray}, \& {Van Speybroeck}}]{vikhlinin2006}
{Vikhlinin}, A., {Kravtsov}, A., {Forman}, W., {Jones}, C., {Markevitch}, M.,
  {Murray}, S.~S., \& {Van Speybroeck}, L. 2006, \apj, 640, 691,
  astro-ph/0507092

\bibitem[{{Voit}(2005)}]{voit2005}
{Voit}, G.~M. 2005, Reviews of Modern Physics, 77, 207, astro-ph/0410173

\bibitem[{{Zwicky}(1933)}]{zwicky1933}
{Zwicky}, F. 1933, Helvetica Physica Acta, 6, 110

\end{thebibliography}
